\documentclass[12pt]{article}

\usepackage{epsf}
\usepackage{graphicx}
\usepackage{amssymb}

\newcommand{\be}{\begin{equation}}
\newcommand{\ee}{\end{equation}}
\newcommand{\bea}{\begin{eqnarray}}
\newcommand{\eea}{\end{eqnarray}}

\newcommand{\cO}{{\cal O}}
\newcommand{\cL}{{\cal L}}

\def\nn{\nonumber}

\def\l{\left}
\def\r{\right}
\def\nn{\nonumber}

\def\beq{\begin{equation}}
\def\eeq{\end{equation}}
\def\barr{\begin{array}}
\def\earr{\end{array}}

\def\ln#1{\log{\l( #1 \r)}}

\def\dfrac#1#2{{\displaystyle {#1 \over #2}}}
\def\dsum{\mathop{\displaystyle \sum }}

\def\simge{\mathrel{\rlap{\raise 0.511ex \hbox{$>$}}{\lower 0.511ex \hbox{$\sim$}}}}
\def\simle{\mathrel{\rlap{\raise 0.511ex \hbox{$<$}}{\lower 0.511ex \hbox{$\sim$}}}} 
\def\slash#1{\setbox0=\hbox{$#1$}\dimen0=\wd0                    
      \setbox1=\hbox{/} \dimen1=\wd1 \ifdim\dimen0>\dimen1
      \rlap{\hbox to \dimen0{\hfil/\hfil}} #1                        \else                                       
      \rlap{\hbox to \dimen1{\hfil$#1$\hfil}}
      /   \fi}                                         

\begin{document}

\thispagestyle{empty}

{\normalsize
\rightline{LPT Orsay, 04-19}
\rightline{RM3-TH/04-3}
\rightline{ROMA-1371/04}
}

\vskip 1.0 true cm 

\begin{center}

{\Large\bf The $K \to \pi$ vector form factor \\ [5pt]
           at zero momentum transfer on the lattice
           \footnote{\bf To appear in Nuclear Physics B.}} \\ [25 pt]

{\sc D. Be\'cirevi\'c${}^1$,  G. Isidori${}^2$, V. Lubicz${}^{3,4}$, 
G. Martinelli${}^5$, \\ F. Mescia${}^{2,3}$, S. Simula${}^4$, 
C. Tarantino${}^{3,4}$}, {\sc G. Villadoro${}^5$} \\ [25 pt]

{\sl ${}^1$Laboratoire de Physique Th\'eorique, Universit\'e Paris Sud, \\
           Centre d'Orsay, F-91405 Orsay-Cedex, France} \\  [3pt]
{\sl ${}^2$INFN, Laboratori Nazionali di Frascati, Via E. Fermi 40, I-00044 Frascati, Italy} \\  [3pt]
{\sl ${}^3$Dip. di Fisica, Universit\`a di Roma Tre, Via della Vasca Navale 84, I-00146 Rome, Italy} \\  [3pt]
{\sl ${}^4$INFN, Sezione di Roma III, Via della Vasca Navale 84, I-00146 Rome, Italy} \\  [3pt]
{\sl ${}^5$Dipartimento di Fisica, Universit\`a di Roma ``La Sapienza'',  \\ 
           and INFN, Sezione di Roma, P.le A. Moro 2, I-00185 Rome, Italy} \\ [30 pt]

{\bf Abstract} 

\end{center} 

\noindent We present a quenched lattice study of the form factors $f_+(q^2)$ 
and $f_0(q^2)$ of the matrix elements $\langle \pi | \bar{s} \gamma_\mu u | 
K \rangle$. We focus on the second-order SU(3)-breaking quantity $[1 - f_+(0)]$, 
which is necessary to extract $|V_{us}|$ from $K_{\ell 3}$ decays. For this 
quantity we show that it is possible to reach the percent precision which is 
the required one for a significant determination of $|V_{us}|$. The leading 
quenched chiral logarithms are corrected for by using analytic 
calculations in quenched chiral perturbation theory. Our final result, 
$f_+^{K^0\pi^-}(0) = 0.960 \pm 0.005_{\rm stat} \pm 0.007_{\rm syst}$, 
where the systematic error does not include the residual quenched 
effects, is in good agreement with the estimate made by Leutwyler and Roos. 
A comparison with other non-lattice computations and the impact of our result 
on the extraction of $|V_{us}|$ are also presented. 

\newpage

\section{Introduction}

The most precise determination of the Cabibbo angle,
or equivalently of the CKM matrix element $|V_{us}|$~\cite{CKM}, is 
obtained from $K \to \pi \ell \nu$ ($K_{\ell3}$) decays.
The key observation which allows to reach a good theoretical 
control on these transitions is the Ademollo-Gatto theorem \cite{ag},
which states that the $K_{\ell3}$ form factors, $f_+(q^2)$ and $f_0(q^2)$, 
at zero four-momentum transfer, are renormalized only by  
terms of at least second order in the breaking of the SU(3) flavor symmetry.
The estimate of these smallish corrections, 
i.e.~of the difference of $f_+(0) = f_0(0)$ from unity, 
is presently the dominant source of theoretical uncertainty 
in the extraction of $|V_{us}|$. 

Chiral perturbation theory (CHPT) provides a natural and powerful tool
to analyse the amount of SU(3) (and isospin) breaking due 
to light quark masses. As shown by Leutwyler 
and Roos \cite{LeuRo}, within CHPT one can perform 
a systematic expansion of the type $f_+(0) = 1 + f_2 + f_4 + \ldots$, 
where $f_n = \cO[{M^n_{K,\pi}}/(4\pi f_\pi)^n]$. 
Because of the Ademollo-Gatto theorem, the first non-trivial 
term in the chiral expansion, 
$f_2$, does not receive contributions of local operators appearing in
the effective theory and can be computed 
unambiguously in terms of $M_K$, $M_\pi$ and $f_\pi$ 
($f_2 = -0.023$ in the $K^0 \to \pi^-$ case \cite{LeuRo}).
The problem of estimating $f_+(0)$ can thus be 
re-expressed as the problem of finding a prediction for
\beq
\Delta f = 1 + f_2 - f_+(0) ~ .
\label{eq:Deltaf}
\eeq
This quantity is difficult to be evaluated since 
it depends on unknown coefficients 
of $\cO(p^n)$ chiral operators, with $n\geq 6$.
Using a general parameterization of the 
SU(3) breaking structure of the pseudoscalar meson 
wave functions, Leutwyler and Roos estimated 
$\Delta f = (0.016 \pm 0.008)$. Very recently, Bijnens 
and Talavera \cite{BT} showed that, in principle, 
the leading contribution to $\Delta f$ could be constrained 
by experimental data on the slope and curvature of $f_0(q^2)$;
however, the required level of experimental precision is far from the 
presently available one.
For the time being we are therefore left with the Leutwyler-Roos result,  
and the large scale dependence of the $\cO(p^6)$ loop calculations 
\cite{BT,post} seems to indicate that its $0.008$ 
error might well be underestimated \cite{CKNRT2}.

The theoretical error on $|V_{us}|$ due to the Leutwyler-Roos
estimate of $f_+(0)$ is already comparable with the present 
experimental uncertainty (see e.g.~Ref.~\cite{CKMBook}).
When the high-statistics $K_{\ell3}$ results 
from KLOE and NA48 will be available, this theoretical  
error will become the dominant source of uncertainty on  $|V_{us}|$.
Given this situation, it is then highly desirable to obtain 
independent estimates of $f_{+}(0)$ at the
$\approx 1$\% level (or below). 
The purpose of the present work is to show that this precision can be achieved
using lattice QCD. 

The strategy adopted in order to reach the challenging goal of a $\approx 1$\%
error, is based on the following three main steps:

\begin{enumerate}

\item[{\bf 1.}]  {\em \underline{Evaluation of the scalar form factor $f_0(q^2)$ at 
      $q^2 = q^2_{\rm max}=(M_K-M_\pi)^2$.}} \\
Applying a method originally proposed in Ref.~\cite{FNAL} to investigate
heavy-light form factors, we extract $f_0(q^2_{\rm max})$ from the relation:
\beq
\frac{ \langle \pi | \bar{s} \gamma_0 u  | K \rangle \langle K | \bar{u} \gamma_0 s | \pi \rangle }{ 
       \langle \pi | \bar{u} \gamma_0 u | \pi \rangle \langle K | \bar{s} \gamma_0 s | K \rangle } =  
\frac{(M_K + M_\pi)^2}{4 M_K M_\pi}\left[ f_0(q^2_{\rm max}; M_K, M_\pi)
\right]^2 ~ ,
\label{eq:fnal}
\eeq 
where all mesons are at rest. The double ratio and the kinematical configuration 
allow to reduce most of the systematic uncertainties and to reach a statistical 
accuracy on $f_0(q_{\rm max}^2 )$ well below $1 \%$.
 
\item[{\bf 2.}] {\em \underline{Extrapolation of $f_0(q_{\rm max}^2)$ to $f_0(0)$.}} \\
By evaluating the slope of the scalar form factor, we extrapolate $f_0$ from $q_{\rm max}^2$ to 
$q^2 = 0$. We note that in order to obtain $f_0(0)$ at the 
percent level the precision required for the slope can be much lower, because it is
possible to choose values of $q_{\rm max}^2$ very close to $q^2 = 0$.

For each set of quark masses we calculate two- and three-point correlation functions
of mesons with various momenta in order to study the $q^2$ dependence of both $f_0(q^2)$ and
$f_+(q^2)$. The latter turns out to be well determined on the lattice, whereas the 
former does not. We improve the precision in the extraction of $f_0(q^2)$ 
by constructing a new suitable double ratio which provides an accurate determination 
of the ratio $f_0(q^2) / f_+(q^2)$. We will define this ratio in Section 4. 
Fitting the $q^2$-dependence of $f_0(q^2)$ with different functional forms, 
we finally extrapolate $f_0(q_{\rm max}^2)$ to $f_0(0)$.
The systematic error induced by this extrapolation, which is strongly reduced by the 
use of small values for $q_{\rm max}^2$, is estimated by the spread of the results 
obtained with different extrapolation functions.

\item[{\bf 3.}] {\em \underline{Subtraction of the leading chiral logs and chiral extrapolation.}} \\
The Ademollo-Gatto theorem holds also within the quenched approximation \cite{cp-q}, 
which has been adopted in this work. The leading $\cO(p^4)$ chiral corrections 
to $f_0(0)$, denoted by $f_2^q$ where the superscript $q$ refers to the quenched 
approximation, are finite and can be computed unambiguously 
in terms of the $\cO(p^2)$ couplings of the quenched CHPT (qCHPT) Lagrangian \cite{bg-q,Sharpe}. 
For these reasons, in order to get rid of some of the quenched artifacts
we define the quantity 
\beq 
R(M_K, M_\pi) \equiv \dfrac{\Delta f}{(\Delta M^2)^2} =
 \dfrac{  1 + f^q_2(M_K, M_\pi) - f_0(0; M_K, M_\pi) }{(\Delta M^2)^2 } ~ , 
  \label{eq:def_R}
\eeq
where $\Delta M^2 \equiv M_K^2 - M_{\pi}^2$ and extrapolate it to the 
physical kaon and pion masses. The ratio $R(M_K, M_\pi)$: i) is finite in the SU(3)-symmetric
limit; ii) does not depend on any subtraction scale; iii) is free from the dominant quenched chiral logs.
We emphasize that the subtraction of $f_2^q$ in Eq.~(\ref{eq:def_R}) does not imply necessarily
a good convergence of (q)CHPT at order $\cO(p^4)$ for the meson masses used in our lattice simulations.
The aim of this subtraction is to define the quantity $\Delta f$ in such a way that its chiral expansion starts 
at order $\cO(p^6)$ independently of the values of the meson masses.
In the presence of sizable local contributions, we expect $R(M_K, M_\pi)$
to have a smooth chiral behavior and to be closer to its unquenched analog than
the SU(3)-breaking quantity $[1 - f_+(0)]$. Extrapolating the values of $R(M_K, M_\pi)$ 
to the physical meson masses, we finally obtain
\beq
\Delta f = R(M_K^{\rm phys}, M_\pi^{\rm phys}) \times [(\Delta M^2)^2]^{\rm phys} = 
(0.017 \pm 0.005_{\rm stat} \pm 0.007_{\rm syst} ) ~ ,
\label{eq:DF}
\eeq
where the systematic error does not include an estimate of quenched 
effects beyond $\cO(p^4)$. Our result (\ref{eq:DF}) is in good agreement with 
the estimate $\Delta f = (0.016 \pm 0.008)$ obtained by Leutwyler and Roos in 
Ref.~\cite{LeuRo}.

The systematic error quoted in Eq.~(\ref{eq:DF}) is mainly due to the uncertainties 
resulting from the functional dependence of the scalar form factor on both $q^2$ and the meson 
masses. This error can be further reduced by using larger lattice volumes (leading to smaller 
lattice momenta) as well as smaller meson masses. In our estimate of $\Delta f$ 
discretization effects start at $\cO(a^2)$ and are also proportional to 
$(m_s - m_u)^2$, as the physical SU(3)-breaking effects. In other words, our result is not
affected by the whole discretization error on the three-point correlation function, but only by its 
smaller SU(3)-breaking part. Discretization errors on $\Delta f$ are estimated to be few percent 
of the physical term, i.e.~well within the systematic uncertainty quoted in Eq.~(\ref{eq:DF}). 
For a more refined estimate of these effects, calculations at different values of the lattice spacing are required. 
Finally, we stress again that the effects of quenching onto the terms of 
$\cO(p^4)$ are not estimated and thus not included in our final systematic 
error.

\end{enumerate}

\noindent Using the unquenched result for $f_2 = - 0.023$ (in the $K^0 \to \pi^-$ case)
\cite{LeuRo}, our final estimate for $f_+^{K^0\pi^-}(0)$ is given by
\bea
f_+^{K^0\pi^-}(0) =  1 + f_2 - \Delta f = 0.960 \pm 0.005_{\rm stat} \pm 0.007_{\rm syst} 
= 0.960 \pm 0.009 ~ .
\eea

\noindent
The plan of the paper is as follows. In Section 2 we introduce the notation
and give some details about the lattice simulation. Section 3 is devoted to the 
extraction of $f_0(q_{\rm max}^2)$ by means of the double ratio method, 
while in Section 4  we study the $q^2$ dependence of the form factors and 
extrapolate the scalar form factor to $q^2 = 0$. The calculation and the subtraction of the 
quenched chiral logarithms as well as the 
extrapolation of $\Delta f$ to the physical masses
is discussed in Section 5. The final estimate of $f_+(0)$, 
its comparison with non-lattice results and the impact on $|V_{us}|$ 
are discussed in Section 6. Finally our conclusions are given in
Section 7.

\section{Notations and lattice details
\label{sec:basics}}

The $K\to\pi$ form factors of the weak vector current 
$V_{\mu} = \bar{s} \gamma_{\mu} u$ are defined by
\beq
\langle \pi^i(p^\prime) | V_{\mu} | K^i(p)  \rangle = C_i
\left[f^i_+(q^2)(p+p^{\prime})_\mu +f^i_-(q^2)(p-p^{\prime})_\mu \right]~, 
\qquad  q^2 = (p-p^\prime)^2~,  
\label{eq:ff}
\eeq
where $C_{i}$ is a Clebsch-Gordan coefficient, equal 
to 1 ($2^{-1/2}$) for neutral (charged) kaons. 
As usual, we express $f^i_- (q^2)$ in terms of the so-called scalar form factor,
\be
f^i_0 (q^2) = f^i_+(q^2)  +  \frac{q^2}{M_K^2 - M_\pi^2}  f^i_- (q^2) \ . 
\ee
By construction $f^i_{0}(0)=f^i_+(0)$ and the differences between $K^0\to \pi^-$ and 
$K^+\to \pi^0$ channels are only due to isospin-breaking effects. In the following we shall
concentrate on the $K^0\to \pi^-$ case and work in the isospin-symmetric 
limit\footnote{For a discussion of isospin breaking effects see the recent review in 
Ref.~\cite{CKMBook}.}, dropping also the superscript $i$ on the form factors.

From Eq.~(\ref{eq:ff}) the form factors can be expressed as linear
combinations of hadronic matrix elements. The latter can be obtained on the lattice 
by calculating two- and three-point correlation functions
\bea
\label{eq:c3pt}
C_\mu^{K \pi} (t_x,t_y,\vec p,\vec{p}^{\,\prime}) & = & \dsum_{\vec x, \vec y} 
\langle O_\pi(t_y,\vec y) ~ \widehat{V}_\mu(t_x,\vec x) ~ O_K^\dagger(0) 
\rangle \, 
e^{-i \vec p \cdot \vec x + i \vec{p}^{\,\prime} \cdot (\vec x - \vec y)}\,\,, \\
C^{K(\pi)} (t,\vec p ) & = & \dsum_{\vec x} 
\langle O_{K(\pi)}(t,\vec x) ~ O_{K(\pi)}^\dagger(0) 
\rangle \, 
e^{-i \vec p \cdot \vec x}\,\,,
\eea
where $\widehat{V}_{\mu}$ is the 
renormalized lattice vector current and 
$O_\pi^{\dagger} = \bar d \gamma_5 u$, 
$O_K^{\dagger} = \bar d \gamma_5 s$ are the operators interpolating $\pi$ and $K$ mesons.
Since we do not consider SU(2)-breaking effects, we always use 
degenerate $u$ and $d$ quarks.

Using the completeness relation and taking $t_x$ and 
$(t_y - t_x)$ large enough, one gets
\bea
C_\mu^{K \pi} (t_x,t_y,\vec p,\vec{p}^{\,\prime}) ~ 
 _{\overrightarrow{\stackrel{\mbox{\tiny $t_x \to \infty$}}{\mbox{\tiny $(t_y - t_x) \to \infty$}}}}
~ \dfrac{\sqrt{Z_K Z_\pi}} {4 E_K E_\pi} 
\langle \pi(p^\prime) | \widehat{V}_\mu
| K(p) \rangle \, e^{- E_K t_x - E_\pi (t_y - t_x)}\,,
\label{eq:c3ptexp}
\eea
\bea
C^{K(\pi)} (t,\vec p (\vec{p}^{\,\prime})) 
& _{\overrightarrow{ {\mbox{\tiny $t \to \infty$}} }} & 
\dfrac{Z_{K(\pi)}}{2 E_{K(\pi)}} 
e^{- E_{K(\pi)} t}\,,
\label{eq:c2ptexp}
\eea
where $E_K = \sqrt{M_K^2 + |\vec p|^2}$, $E_{\pi} = \sqrt{M_{\pi}^2 + 
|\vec{p}^{\,\prime}|^2}$ and $\sqrt{Z_{K(\pi)}} = \langle 0 | O_{K(\pi)}(0) | K(\pi)
\rangle$. Then it follows
\bea
\dfrac{C_\mu^{K \pi} (t_x,t_y,\vec p,\vec{p}^{\,\prime})}{C^K (t_x,\vec p )
~ C^{\pi} (t_y - t_x, \vec{p}^{\,\prime})} \left.
 _{\overrightarrow{\stackrel{\mbox{\tiny $t_x \to \infty$}}{\mbox{\tiny $(t_y - t_x) \to \infty$}}}}
\right. \dfrac{\langle \pi(p^\prime) | \widehat{V}_\mu | K(p) \rangle }
{\sqrt{Z_K Z_\pi}}\,.
\label{eq:standard}
\eea
Consequently the hadronic matrix elements $\langle \pi(p^\prime) | \widehat{V}_\mu
| K(p) \rangle$ can be obtained from the plateaux of the l.h.s.~of 
Eq.~(\ref{eq:standard}), once $Z_K$ and $Z_\pi$ are separately extracted from 
the large-time behavior of the two-point correlators (\ref{eq:c2ptexp}).

The procedure described above is the standard one to calculate form factors on the lattice.
In this way, however, it is very hard to reach the percent level precision required for the
present calculation. In the next Section we describe a very efficient procedure to get both 
the scalar form factor at $q^2 = q_{\rm max}^2$ and the ratio $f_0(q^2) / f_+(q^2)$ with quite 
small statistical fluctuations.

Numerical data have been obtained in the quenched approximation on a $24^3 \times 56$
lattice, by using the plaquette gluon action at $\beta = 6.2$. In order
to remove leading discretization effects, the non-perturbatively 
$\mathcal{O}(a)$-improved Wilson action and fermionic bilinear operators are
employed~\cite{improvement}. In case of the weak vector current one has
\be
\widehat{V}^\mu = Z_V \left(1 + b_V \dfrac{a m_s + a m_\ell}{2} \right) 
\left( \bar{s} \gamma^{\mu} u + c_V \partial_\nu ~ \bar{s} \sigma^{\mu \nu} u 
\right) \,,
\label{eq:vtilde}
\ee
where $Z_V$ is the vector renormalization constant, $b_V$ and $c_V$ are $O(a)$-improvement 
coefficients and the subscript $\ell$ refers to the light $u$ (or $d$) quark.

We performed a first run by generating $230$ gauge field configurations 
and choosing quark masses corresponding to four values of the hopping 
parameters, namely $k \in \{ 0.13390, 0.13440, 0.13490, 0.13520 \}$.
Using $K$ and $\pi$ mesons with quark content ($k_s k_\ell$) and ($k_\ell k_\ell$) 
respectively, twelve different $K \to \pi$ correlators ($C_\mu^{K \pi}$) have been computed, using 
both $k_s < k_\ell$ and $k_s > k_\ell$, corresponding to the cases in which the 
kaon(pion) is heavier than the pion(kaon). In addition, using the same combinations
of quark masses, also the three-point $\pi \to K$ correlations ($C_\mu^{\pi K}$) have been
calculated. Finally, twelve non-degenerate $K \to K$ and four degenerate 
$\pi \to \pi$ three-point functions have been evaluated.

In order to increase the number of different combinations of quark masses, we generated a second run 
of $230$ independent gauge configurations with quark propagators computed at three new values of the
hopping parameter, namely $k \in \{ 0.13385, 0.13433, 0.13501 \}$. In this way six 
different combinations of meson masses have been added to the first run, obtaining a total 
of 18 independent sets of pseudoscalar meson masses.

The simulated quark masses are approximately in the range ($0.5 \div 2$) $\times ~ m_s$, 
where $m_s$ is the strange quark mass, and correspond to $K$ and $\pi$ meson masses in the 
interval $\approx 0.5 \div 1$ GeV. Though the simulated meson masses are larger than the physical
ones, the corresponding values of $q_{\rm max}^2 = (M_K - M_\pi)^2$ are taken as close as 
possible to $q^2 = 0$ (see Table \ref{tab:f0qmax} in the next Section).

As for the critical hopping parameter, we find $k_c = 0.135820(2)$ 
using the axial Ward identity. The physical values of meson masses in lattice units
have been obtained using the method of ``lattice physical planes" \cite{plane}, in which the ratios 
$M_K / M_{K^*}$ and $M_\pi / M_\rho$ are fixed to their experimental values. 
In this way we find 
\bea
[a M_K]^{\rm phys} = 0.189(2) \,, & [a M_\pi]^{\rm phys} = 0.0536(7)\,.
\label{eq:physmasses}
\eea

To improve the statistics, two- and three-point correlation functions have been averaged with 
respect to spatial rotation, parity and charge conjugation transformations. For the same 
reason we have chosen $t_y = T/2$ in the three-point correlators, which allows to average 
the latter between the left and right halves of the lattice. Finally three-point correlation 
functions have been computed for the $10$ different combinations of momenta 
$(\vec p, \vec{p}^{\,\prime})$ listed in Table~\ref{tab:pq}.

\begin{table} [htb]
\begin{center}
\begin{tabular}{||c||c|c|c|c||c|c|c|c|c||}
\hline 
$a \vec p$ & {\footnotesize $(0,0,0)$} & {\footnotesize $(1,0,0)$}  & 
{\footnotesize $(1,1,0)$} & {\footnotesize $(1,1,1)$} & 
{\footnotesize $(0,0,0)$} & {\footnotesize $(\pm 1,0,0)$} & 
{\footnotesize $(1,1,0)$} & {\footnotesize $(0,1,1)$} & 
{\footnotesize $(0,1,0)$}\\
\hline 
$a \vec{p}^{\,\prime}$ & {\footnotesize $(0,0,0)$} & {\footnotesize $(0,0,0)$} & 
{\footnotesize $(0,0,0)$} & {\footnotesize $(0,0,0)$} & 
{\footnotesize $(1,0,0)$} & {\footnotesize $(1,0,0)$} & 
{\footnotesize $(1,0,0)$} & {\footnotesize $(1,0,0)$} & 
{\footnotesize $(1,0,0)$}\\
\hline
\end{tabular}
\end{center}
\caption{\it Momentum configurations in units of $2 \pi / L = \pi / 12$ of 
three-point correlators with $\vec p$ ($\vec{p}^{\,\prime}$) being the kaon(pion)
spatial momentum.} 
\label{tab:pq}
\end{table}

The statistical errors are evaluated using the jacknife procedure for each run. To combine 
the results of a given observable obtained in the two runs, we have used a Monte Carlo 
procedure to generate values of the observable which are normally distributed around their 
averages with widths given by the jacknife errors. Such a procedure is adopted throughout
this paper, including the chiral extrapolation of $f_0(0)$ described in Section~\ref{sec:Df}.

\section{Calculation of $f_0(q^2_{\rm max})$}
\label{sec:f0q2max}

\begin{figure}[t]

\begin{center}

\epsfxsize=11.5cm \epsffile{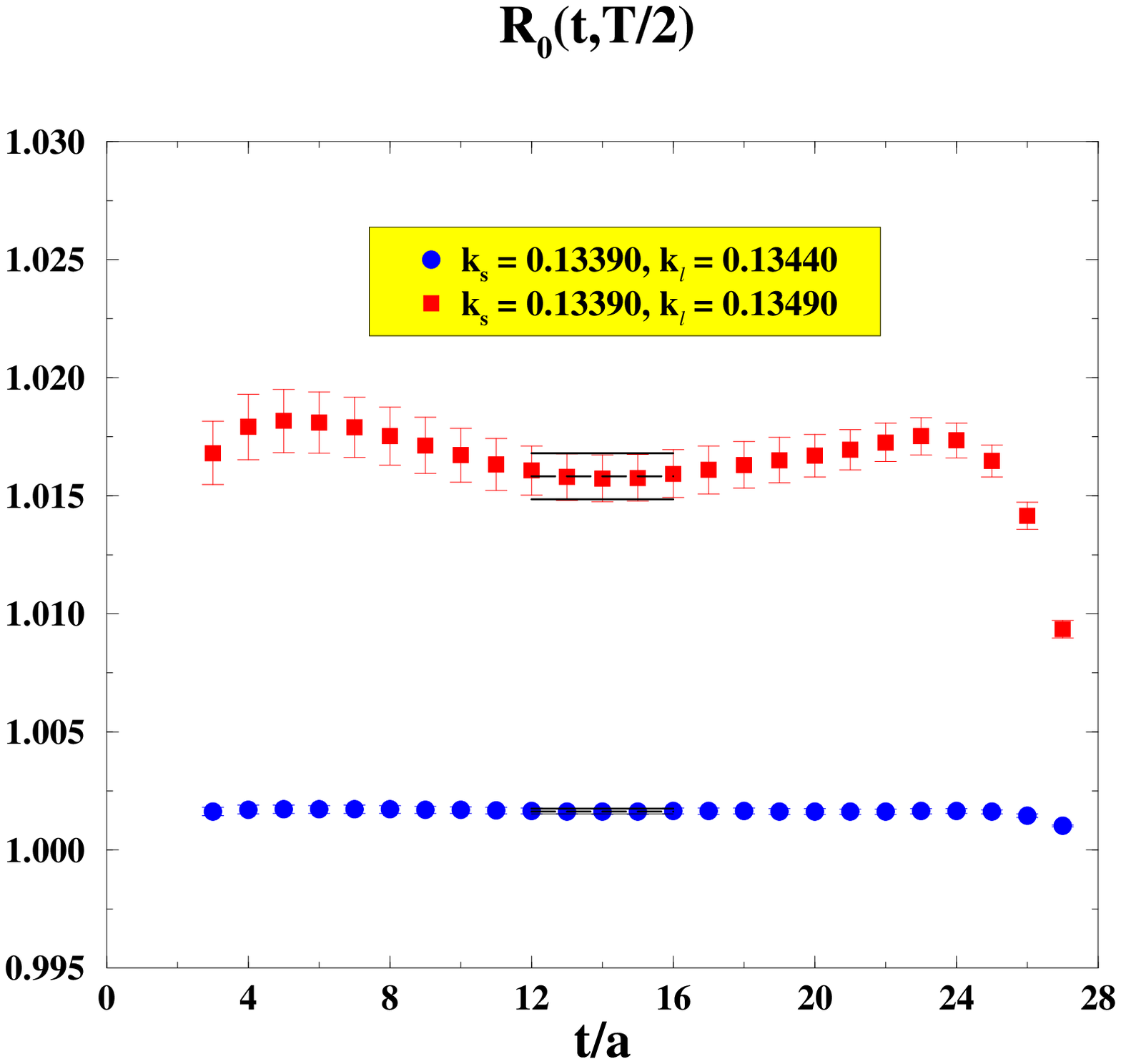}

\end{center}

\caption{\it Time dependence of the double ratio (\protect\ref{eq:fnal1}), averaged between the
two halves of the lattice, for two sets of quark masses, given in the legend. The plateaux are 
taken from $t/a = 12$ to $t/a = 16$.}

\label{fig:R0plateau}

\end{figure}

Following a procedure originally proposed in Ref.~\cite{FNAL} to study the
heavy-light form factors, the scalar form factor has been 
calculated very efficiently at $q^2 = q^2_{\rm max} = (M_K - M_\pi)^2$ (i.e. 
$\vec p = \vec{p}^{\,\prime} = \vec q = 0$) from the double ratio of three-point
correlation functions with both mesons at rest:
\be
R_0(t_x, t_y) \equiv \dfrac{C_0^{K \pi}(t_x,t_y,\vec 0,\vec 0) \, C_0^{\pi K}
(t_x,t_y,\vec 0,\vec 0)}
{C_0^{K K}(t_x,t_y,\vec 0,\vec 0) \, C_0^{\pi \pi}(t_x,t_y,\vec 0,\vec 0)}
\,.
\label{eq:fnal1}
\ee
When the vector current and the two interpolating fields are separated far 
enough from each other, the contribution of the ground states dominates, yielding
\be 
R_0(t_x, t_y) _{\overrightarrow{\stackrel{\mbox{\tiny $t_x \to \infty$}}{\mbox{\tiny $(t_y - t_x) \to \infty$}}}} 
\dfrac{\langle \pi | \bar{s} \gamma_0 u | K \rangle \, 
\langle K | \bar{u} \gamma_0 s | \pi \rangle}{\langle K | \bar{s} \gamma_0 s | K \rangle \, 
\langle \pi | \bar{u} \gamma_0 u | \pi \rangle} 
= [f_0(q^2_{\rm max})]^2 \,\dfrac{(M_K + M_\pi)^2}{4 M_K M_\pi}\,.
\label{eq:fnal2}
\ee
The quality of the plateau for the double ratio can be appreciated by looking at
Fig.~\ref{fig:R0plateau}.

There are several crucial advantages in the use of the double ratio (\ref{eq:fnal1}). 
First, there is a large reduction of statistical uncertainties, because 
fluctuations in the  numerator and the denominator are highly correlated. Second,
the matrix elements of the meson sources, $\sqrt{Z_K}$ and $\sqrt{Z_{\pi}}$, appearing in 
Eq.~(\ref{eq:c3ptexp}), cancel between numerator and denominator. Third, the double ratio 
is independent from the improved renormalization constant $Z_V [1 + b_V (a m_s + a
m_\ell)/2]$ as well as from the improvement coefficient $c_V$ 
[see Eq.~(\ref{eq:vtilde})]~\footnote{The latter feature holds only when $\vec{q} = 0$ and 
the time component of the vector current is used.}. 
Therefore the knowledge of $Z_V$, $b_V$ and $c_V$ is not necessary to compute $R_0$ 
and this ratio is automatically improved at order $a^2$.
Finally, the double ratio is equal to unity in the SU(3)-symmetric limit at all orders 
in the lattice spacing $a$. Thus the deviation of $R_0$ from unity depends on the physical SU(3)
breaking effects on $f_0(q_{\rm max}^2)$ as well as on discretization errors, which are 
at least of order $a^2 (m_s - m_\ell)^2$ (see Section \ref{subsec:discre}). 
A similar consideration applies also to the quenching error, because the double ratio 
$R_0(t_x, t_y)$ is correctly normalized to unity in the SU(3)-symmetric limit also in 
the quenched approximation.

\begin{figure}[htb]

\begin{center}

\parbox{7.75cm}{\epsfxsize=7.70cm 
\epsffile{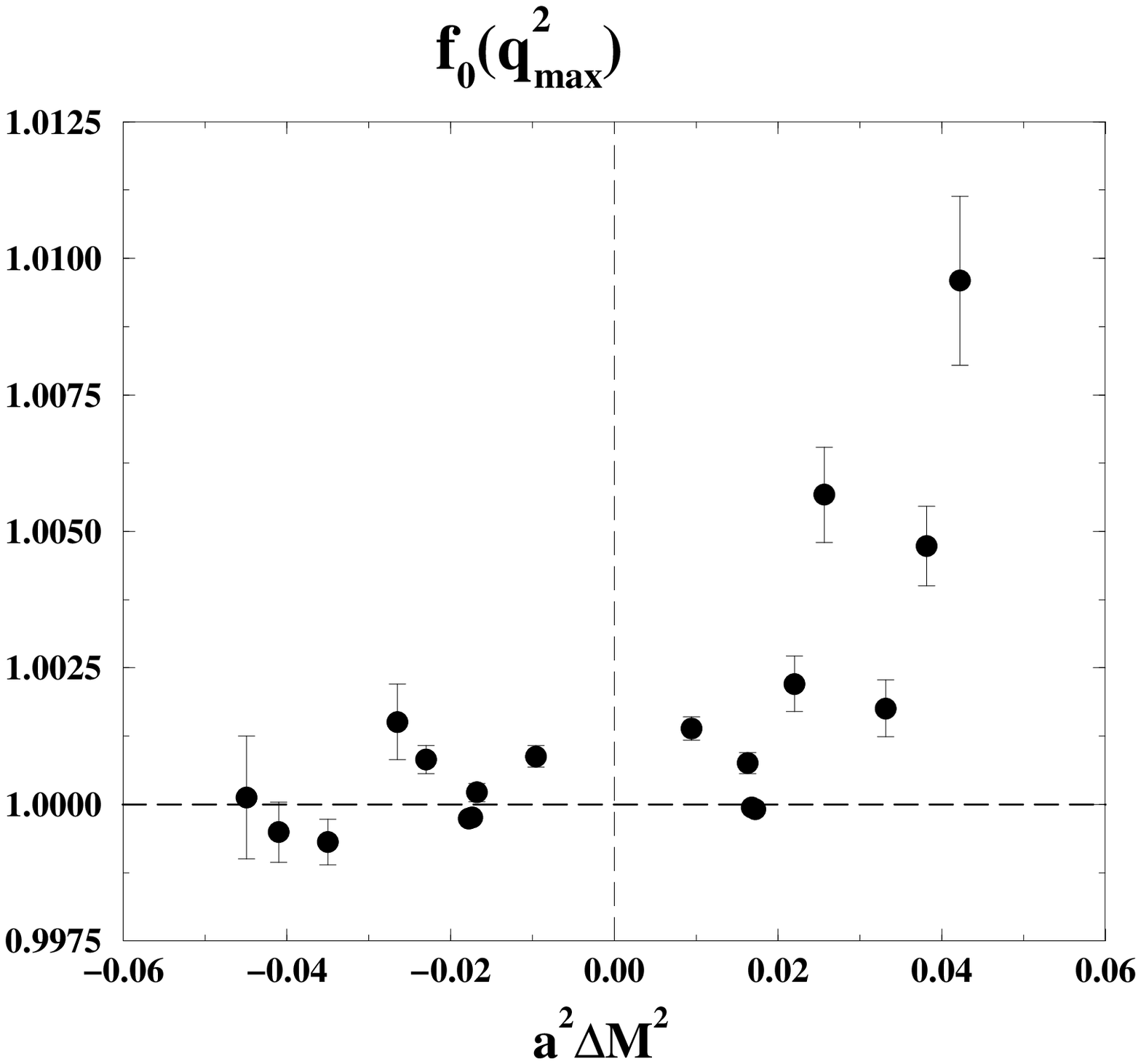}} ~~
\parbox{7.75cm}{\epsfxsize=7.70cm 
\epsffile{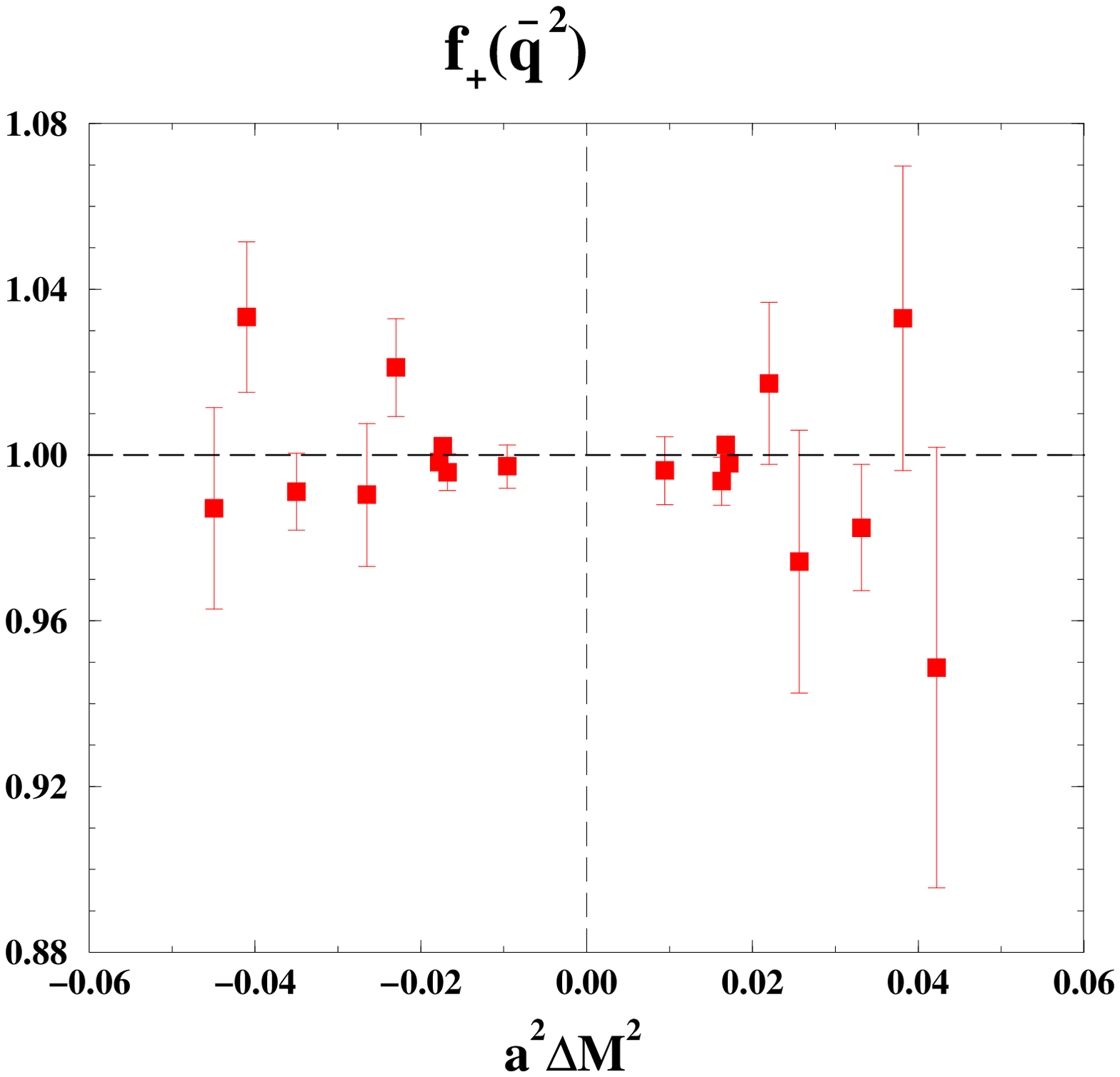}}

\end{center}

\caption{\it Values of $f_0(q_{\rm max}^2)$ (left) and $f_+(\bar{q}^2)$ (right) as a function of 
the SU(3)-breaking parameter $a^2 \Delta M^2 = a^2 M_K^2 - a^2 M_{\pi}^2$. The dependence on 
$q_{\rm max}^2$ and $\bar{q}^2$ is not explicitly shown in the plots.}

\label{fig:f0fpqmax}

\end{figure}

Having chosen $t_y = T/2$, the three-point correlation functions are symmetric between 
the two halves $0 < t_x < T/2$ and $T/2 < t_x < T$. Statistical fluctuations, however,
are independent in the two halves of the lattice. The best precision is reached when 
the double ratio is constructed in the two 
halves separately and then averaged. We obtain values of $f_0(q^2_{\rm max})$ 
with an uncertainty smaller than $0.1 \%$, as it can be seen from Table~\ref{tab:f0qmax} and
Fig.~\ref{fig:f0fpqmax} (left).

\begin{table}[htb]

\vspace{1cm}

\begin{center}
\begin{tabular}{||c||l|l|l||}
\hline 
$k_s - k_\ell$ & $~~~~ a^2 \Delta M^2$ & $ ~~ a^2 q_{\rm max}^2$ & $ ~~ f_0(q_{\rm max}^2)$ \\ \hline \hline
 $0.13390-0.13440 $ & $+0.01727(12)$ & $0.000737(10) $ & $0.99991(6) $\\ \hline
 $0.13390-0.13490 $ & $+0.03316(26)$ & $0.00365(6) $ & $1.0018(5) $\\ \hline
 $0.13390-0.13520 $ & $+0.04225(34)$ & $0.00745(15) $ & $1.010(2) $\\ \hline
 $0.13440-0.13390 $ & $-0.01777(12)$ & $0.000665(8) $ & $0.99974(6) $\\ \hline
 $0.13440-0.13490 $ & $+0.01634(15)$ & $0.000988(21) $ & $1.0008(2) $\\ \hline
 $0.13440-0.13520 $ & $+0.02565(23)$ & $0.00312(8) $ & $1.0057(9) $\\ \hline
 $0.13490-0.13390 $ & $-0.03496(30)$ & $0.00279(5) $ & $0.9993(4) $\\ \hline
 $0.13490-0.13440 $ & $-0.01674(17)$ & $0.000832(18) $ & $1.0002(2) $\\ \hline
 $0.13490-0.13520 $ & $+0.00948(10)$ & $0.000497(15) $ & $1.0014(2) $\\ \hline
 $0.13520-0.13390 $ & $-0.04491(46)$ & $0.00484(11) $ & $1.000(1) $\\ \hline
 $0.13520-0.13440 $ & $-0.02646(31)$ & $0.00222(6) $ & $1.0015(7) $\\ \hline
 $0.13520-0.13490 $ & $-0.00955(12)$ & $0.000416(13) $ & $1.0009(2) $\\ \hline \hline
 $0.13385-0.13433 $ & $+0.01682(7)$ & $0.000669(6) $ & $0.99995(5) $\\ \hline
 $0.13385-0.13501 $ & $+0.03821(24)$ & $0.00519(6) $ & $1.0047(7) $\\ \hline
 $0.13433-0.13385 $ & $-0.01731(7)$ & $0.000610(6) $ & $0.99977(5) $\\ \hline
 $0.13433-0.13501 $ & $+0.02205(15)$ & $0.00193(3) $ & $1.0022(5) $\\ \hline
 $0.13501-0.13385 $ & $-0.04093(22)$ & $0.00380(5) $ & $0.9995(6) $\\ \hline
 $0.13501-0.13433 $ & $-0.02295(15)$ & $0.00154(2) $ & $1.0008(3) $\\ \hline

 \hline

 \end{tabular}
 
 \end{center}
 
 \caption{\it Values of the hopping parameters $k_s$ and $k_\ell$, $a^2 \Delta M^2$,
 $a^2 q_{\rm max}^2$ and $f_0(q_{\rm max}^2)$ obtained with the double ratio method 
 [see Eq.~(\ref{eq:fnal2})].}
 
 \label{tab:f0qmax}
 
 \end{table}
 
By replacing in Eq.~(\ref{eq:fnal1}) the time component of the vector current with the spatial 
ones and setting $a \vec p = a \vec{p}^{\,\prime} = \pi/12 ~ (1,0,0)$ (the minimum non-zero 
momentum allowed on our lattice), also the form factor $f_+$ can 
be extracted from the double ratio
\be
\dfrac{C_1^{K \pi}(t_x,t_y,\vec p,\vec p) \, C_1^{\pi K}
(t_x,t_y,\vec p,\vec p)}
{C_1^{K K}(t_x,t_y,\vec p,\vec p) \, C_1^{\pi \pi}(t_x,t_y,\vec p,\vec p)} \left.
 _{\overrightarrow{\stackrel{\mbox{\tiny $t_x \to \infty$}}{\mbox{\tiny $(t_y - t_x) \to \infty$}}}} \right.
[f_+(\bar{q}^2)]^2
\,.
\label{eq:fpfnal}
\ee
where $\bar{q}^2 = \left(\sqrt{M_K^2 + |\vec{p}|^2} - \sqrt{M_{\pi}^2 + |\vec{p}|^2} \right)^2 < q_{\rm max}^2$.
The uncertainty obtained in this case, however, is about twenty times
larger than the one obtained for $f_0(q^2_{\rm max})$, as shown in Fig.~\ref{fig:f0fpqmax}.

\section{Momentum dependence of the form factors and extrapolation to $q^2 = 0$}
\label{sec:extrq20}

In this Section we perform the extrapolation of the scalar form factor from $q_{\rm max}^2$
to $q^2 = 0$. To this end we need to evaluate the slope of $f_0$, which in turn means to study
the $q^2$-dependence of the scalar form factor. We stress that, in order to obtain $f_0(0)$ at 
the percent level, the precision required for the slope can be much lower, since the values 
of $q_{\rm max}^2$ used in our lattice calculations are quite close to $q^2 = 0$ (see 
Table \ref{tab:f0qmax}). We find indeed that a $\simeq 30 \%$ precision is enough for our 
purposes.

The form factors $f_0(q^2)$ and $f_+(q^2)$ can be expressed as linear combinations of the 
matrix elements $\langle \pi | \hat{V}_{\mu} | K \rangle$. Their $q^2$-dependence is obtained 
by studying these matrix elements determined according to Eq.~(\ref{eq:standard}) for the set of 
momenta listed in Table~\ref{tab:pq}, using 
both time and spatial components of the vector current.

The renormalization constant $Z_V$ of the lattice vector current and the improvement coefficients
$b_V$ and $c_V$ are needed in the calculation. In order to minimize the statistical
fluctuations we have calculated $Z_V$ and $b_V$ on the same sets of gauge configurations 
and combinations of quark masses used in the study of the form factors.

\subsection{Evaluation of $Z_V$ and $b_V$}

The vector current renormalization constant, $Z_V$, and the improvement coefficient, $b_V$,
can be extracted from the following relation
\be
\dfrac{1}{2} \dfrac{C^{K(\pi)}(T/2, \vec 0)}{C_0^{KK(\pi \pi)}(t_x, T/2, \vec 0, \vec 0)}
= Z_V (1 + b_V a m_q) + \mathcal{O}(a^2) \,,
\label{eq:zv}
\ee
where $a m_q = (1 / k_s - 1 / k_c) /2$ for $K \to K$ transitions and $a m_q = (1 / k_\ell - 1 / k_c) /2$ 
for $\pi \to \pi$ ones. The results obtained for the l.h.s.~of Eq.~(\ref{eq:zv}) 
show a clear linear dependence on the quark mass $a m_q$. 
A linear fit yields
\bea
Z_V = 0.7876 \pm 0.0002\,, & b_V = 1.393 \pm 0.006 \,,
\label{eq:zvbv}
\eea
in excellent agreement with the findings of Refs.~\cite{Bhattacharya}-\cite{Becirevic:2004ny}. In
what follows we use the values of $Z_V$ and $b_V$ given in Eq.~(\ref{eq:zvbv}) and adopt for 
the improvement coefficient $c_V$ the non-perturbative value $c_V = -0.09$ from 
Ref.~\cite{Bhattacharya}\footnote{We note that the scalar form factor $f_0(q^2)$ being proportional 
to the matrix element of $\partial^{\mu} \widehat{V}^{\mu}$ is independent of $c_V$ 
[see Eq.~(\ref{eq:vtilde})]. The choice of $c_V$ affects only the determination of the form 
factor $f_+(q^2)$ and its numerical impact turns out to be smaller than the statistical uncertainty.}.

\subsection{$q^2$-dependence of $f_0$ and $f_+$}

\begin{figure}[htb]

\begin{center}

\parbox{7.75cm}{\epsfxsize=7.70cm 
\epsffile{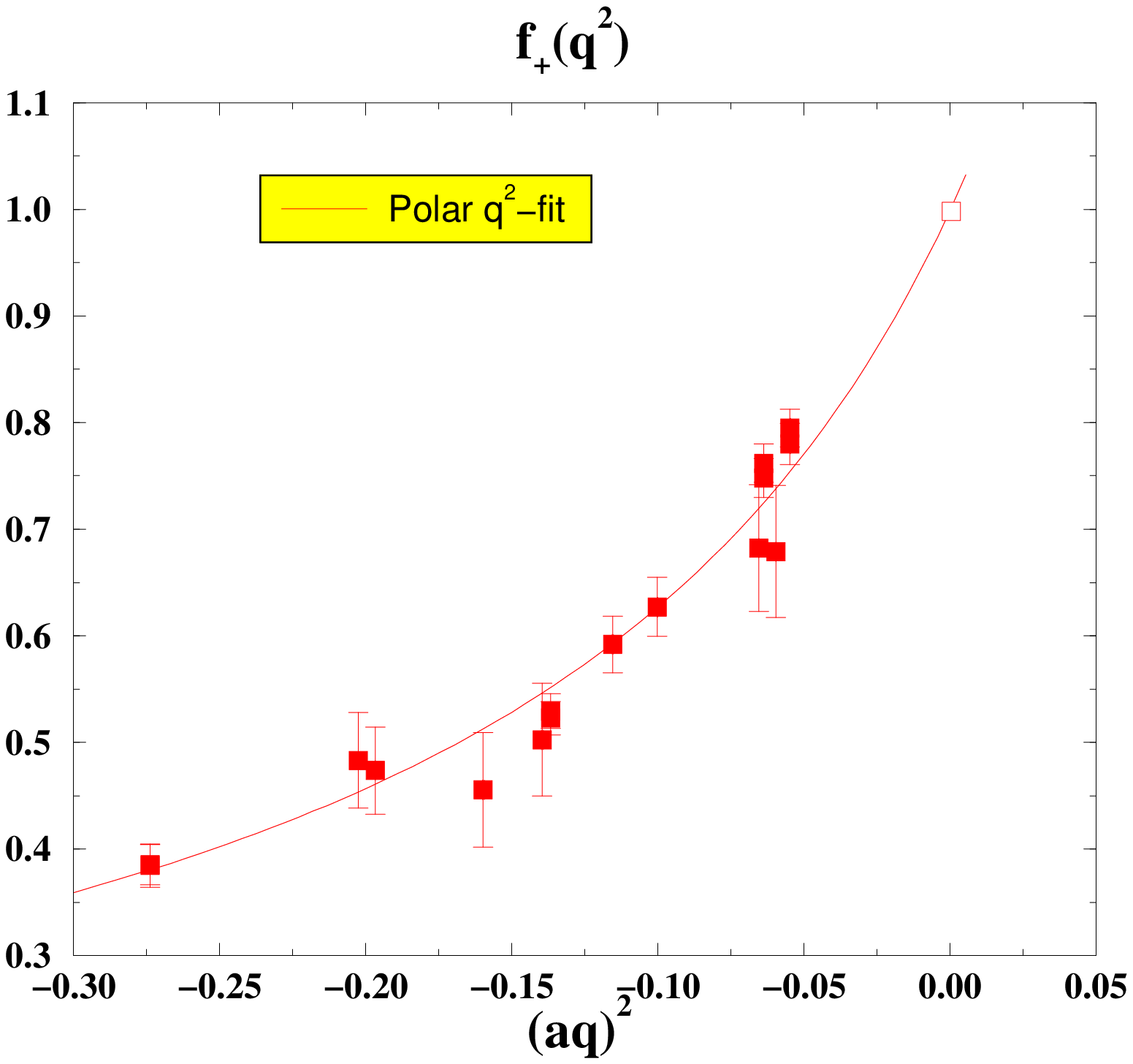}} ~~
\parbox{7.75cm}{\epsfxsize=7.70cm 
\epsffile{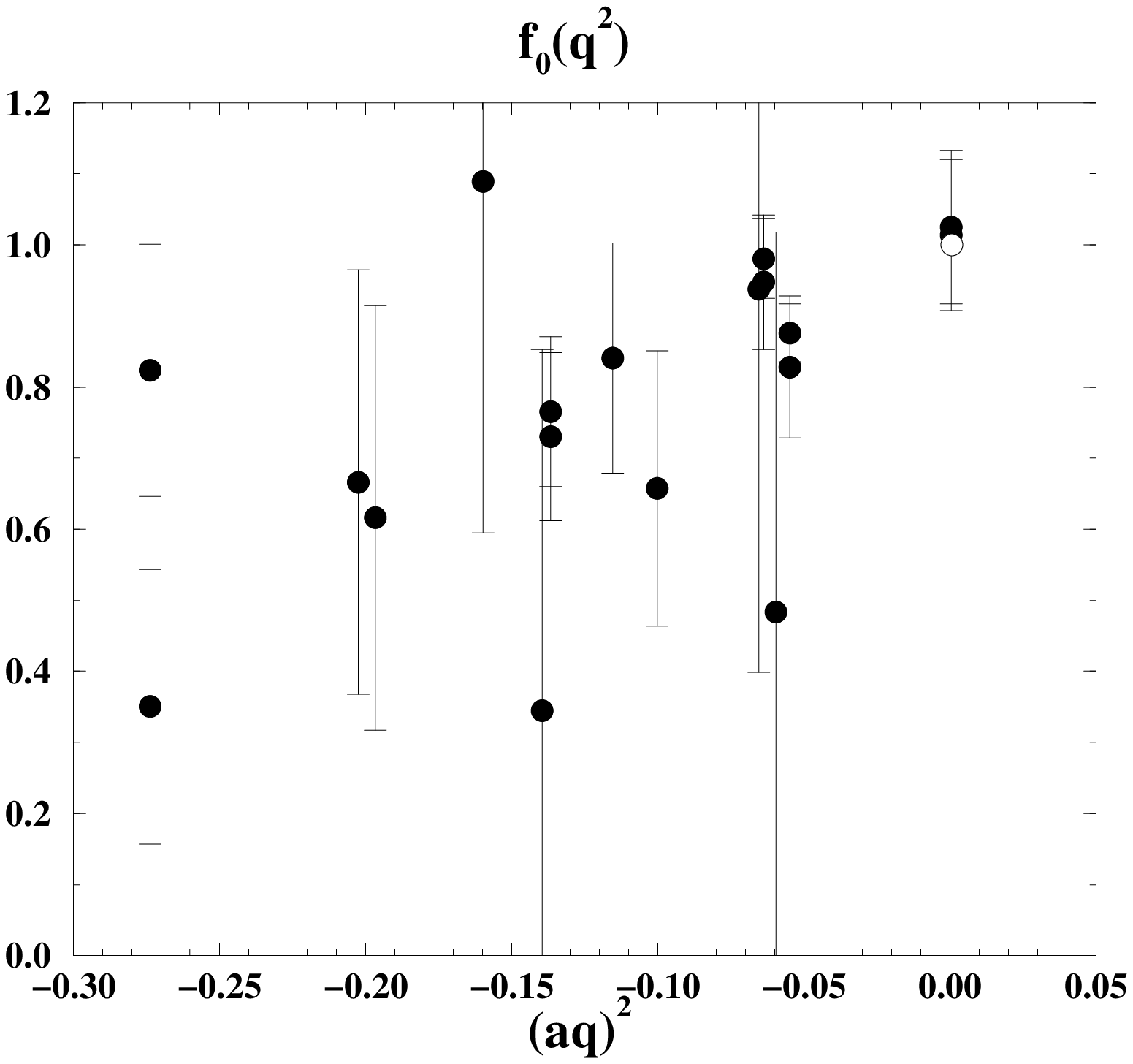}}

\end{center}

\caption{\it Values of the form factors $f_+(q^2)$ (left) and $f_0(q^2)$ (right) as a function of 
the four-momentum square $(a q)^2$, for the quark mass combination 
$k_s = 0.13390$ and $k_\ell = 0.13440$. Open markers represent the results obtained with the 
double ratio method; the errors on these points are smaller than the size of the markers. In the 
left picture the solid line is the pole-dominance fit (\protect\ref{eq:fpfit}).}

\label{fig:f0fp}

\end{figure}

The form factors $f_0(q^2)$ and $f_+(q^2)$ are shown in Fig.~\ref{fig:f0fp} as a function
of $q^2$. Note that, although we considered only ten independent combinations of meson momenta 
(see Table \ref{tab:pq}), we have computed both $K \to \pi$ and $\pi \to K$ amplitudes, obtaining 
in this way twenty different values of $q^2$. The form factor $f_+(q^2)$ is rather well determined 
with a statistical error of $\simeq 5 \div 20 \%$, whereas for the scalar form factor the 
uncertainties turn out to be about $5$ times larger. 

Our results for $f_+(q^2)$ are very well described by a pole-dominance fit, viz.
\bea
f_+(q^2) & = & \dfrac{f_+(0)}{1 - \lambda_+ \, q^2} \,,
\label{eq:fpfit}
\eea
as illustrated by the solid line in Fig.~\ref{fig:f0fp}. The values obtained for 
the slope $\lambda_+$ agree well with the inverse of the $K^*$-meson mass square 
for each combination of the simulated quark masses. A simple linear extrapolation in 
terms of the quark masses to the physical values yields $\lambda_+ =  0.026 \pm 0.002$ 
in units of the physical value of $M_{\pi^+}^2$, which is consistent with the PDG value 
$\lambda_+ =  0.028 \pm 0.002$ \cite{PDG}\footnote{Since experiments 
cover a narrow region of values of $q^2$, the PDG value is obtained assuming 
a linear $q^2$-dependence of the data.} as well as with the recent 
measurement from KTeV $\lambda_+ =  0.02502 \pm 0.00037$ 
\cite{KTEV_slope}, obtained using a pole parameterization.
The polar fit (\ref{eq:fpfit}) also provides the value of $f_+(0)$. Due to the uncertainties in 
the determination of $f_+(q^2)$, however, the values obtained for $f_+(0)$ have statistical 
errors well above $1 \%$, which do not allow to investigate SU(3)-breaking effects. 
The only way to keep these errors below $1 \%$ is to use the 
high-precision results obtained for $f_0(q_{\rm max}^2)$.
This in turn requires a drastic improvement in the evaluation of the scalar
form factor as a function of $q^2$. To this end, we have considered an alternative procedure
based on the introduction of suitable double ratios, namely
\be
R_i(t_x, t_y) = \dfrac{C_i^{K \pi}(t_x,t_y,\vec p,\vec{p}^{\,\prime})}{C_0^{K \pi}(t_x,t_y,\vec p,\vec{p}^{\,\prime})}
 ~ \dfrac{C_0^{K K}(t_x,t_y,\vec p,\vec{p}^{\,\prime})}{C_i^{K K}(t_x,t_y,\vec p,\vec{p}^{\,\prime})} ~ ,
\label{eq:rappd}
\ee
from which a determination of the ratio of the form factors $f_0(q^2) / f_+(q^2)$ can be obtained.
The advantages of the double ratios (\ref{eq:rappd}) are similar to those already pointed out
for the double ratio (\ref{eq:fnal1}), namely: i) a large reduction of statistical fluctuations; 
ii) the independence of the renormalization constant $Z_V$ and the improvement coefficient $b_V$,
and iii) $R_i \to 1$ in the SU(3)-symmetric limit. We stress that the introduction 
of the matrix elements of degenerate mesons in Eq.~(\ref{eq:rappd}) is crucial to largely reduce 
statistical fluctuations, because it compensates the different fluctuations of the matrix 
elements of the spatial and time components of the weak current.

\begin{figure}[htb]

\begin{center}

\epsfxsize=11.5cm \epsffile{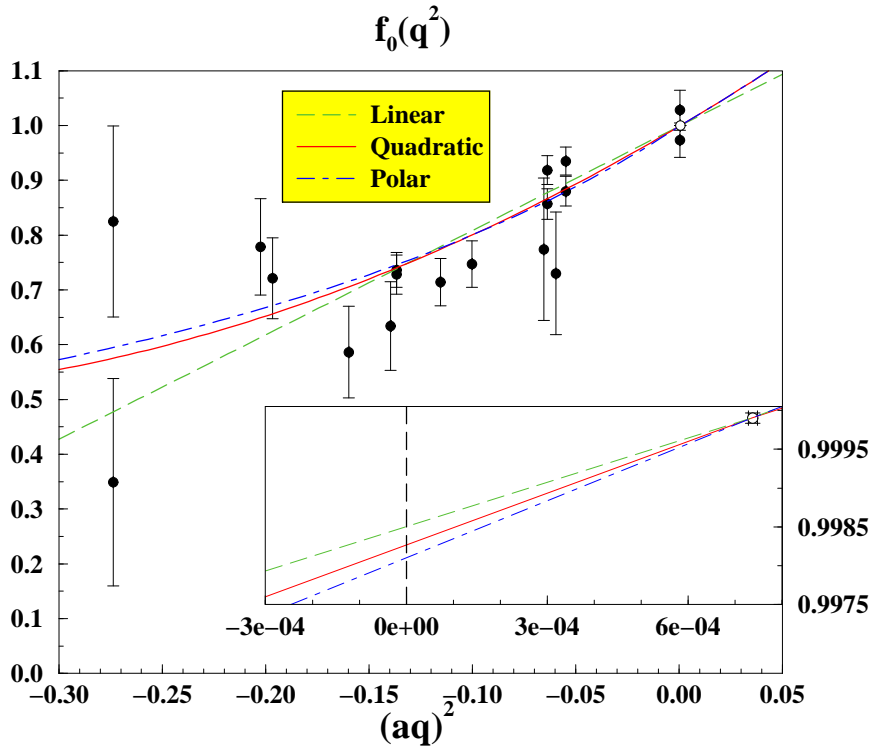} 

\end{center}

\caption{\it The form factor $f_0(q^2)$ obtained from the double ratios
(\ref{eq:rappd}) for $q^2 < q_{\rm max}^2$ (full dots) and from the double ratio (\ref{eq:fnal1}) at
$q^2 = q_{\rm max}^2$ (open dot), for the quark mass combination $k_s = 0.13390$ and $k_\ell = 0.13440$. 
The dot-dashed, dashed and solid lines correspond to the polar, linear and
quadratic fits given in Eqs.~(\ref{eq:polar}-\ref{eq:quadratic}), respectively. The inset is an 
enlargement of the region around $q^2 = 0$.}

\label{fig:f0fit}

\end{figure}

By denoting with $\overline{R}_i(q^2)$ the plateaux of the double ratios
$R_i(t_x, t_y)$ at enough large time distances, the ratio $f_0(q^2)/f_+(q^2)$ is given by
\be
\dfrac{f_0(q^2)}{f_+(q^2)}  =  1 + \dfrac{q^2}{M_K^2 - M_\pi^2}\, 
\dfrac{-(E_K + E_K^\prime)\, (p + p^\prime)_i + (E_K +
E_\pi) (p + p^\prime)_i \, \overline{R}_i(q^2)}
{(E_K + E_K^\prime)\, (p - p^\prime)_i - (E_K -
E_\pi) (p + p^\prime)_i \, \overline{R}_i(q^2)}\,,
\label{eq:f0sfp}
\ee
where $E_K^\prime \equiv \sqrt{M_K^2 + |\vec{p}^{\,\prime}|^2}$. The $q^2$-behavior of 
$f_0(q^2)$ can be then investigated after multiplying the ratio $f_0(q^2) / f_+(q^2)$, 
obtained from Eq.~(\ref{eq:f0sfp}), by the values of $f_+(q^2)$ obtained with 
the standard procedure. The statistical uncertainties on $\overline{R}_i(q^2)$ and $f_0(q^2)$ 
turn out to be $\simeq 1 \div 20 \%$ and $\simeq 5 \div 20 \%$, respectively. The quality of
the results for $f_0(q^2)$, obtained in this way, is illustrated in Fig.~\ref{fig:f0fit} for one of the 
eighteen combinations of quark masses. These results can be directly compared with those 
obtained by using the standard method and shown in Fig.~\ref{fig:f0fp} (right).
 
In order to extrapolate the scalar form factor to zero-momentum transfer we have 
considered three different possibilities, namely a polar, a linear and a quadratic fit:
\bea
f_0(q^2) & = & f_0^{(pol.)}(0) / (1 - \lambda_0^{(pol.)} \, q^2) \,, \label{eq:polar} \\
f_0(q^2) & = & f_0^{(lin.)}(0) \cdot (1 + \lambda_0^{(lin.)} \, q^2) \,, \label{eq:linear} \\
f_0(q^2) & = & f_0^{(quad.)}(0) \cdot (1 + \lambda_0^{(quad.)} \, q^2 + c_0 \, q^4) \,.
\label{eq:quadratic}
\eea
These fits are shown in Fig.~\ref{fig:f0fit}.

The polar, linear and quadratic forms fit our results for all quark masses with comparable 
$\chi^2$ values. The three fits provide both $f_0(0)$ and the slope $\lambda_0$ with
values consistent with each other within the statistical uncertainties. Although tiny, 
the differences found for $f_0(0)$ are appreciable with respect to the size of the
physical SU(3)-breaking effects. Consequently, for each combination of quark 
masses we consider all three extrapolated values of $f_0(0)$ and treat the difference 
as a systematic uncertainty in the rest of our analysis. The results obtained for $f_0(0)$ 
and the slope $\lambda_0$ are collected in Table \ref{tab:f0(0)}.

\begin{table}[htb]

\vspace{1cm}

\begin{center}
\begin{tabular}{||c||l|l||l|l||l|l||}
\hline 
$k_s - k_\ell$ & $f_0^{(pol.)}(0)$ & $\lambda_0^{(pol.)}$ & $f_0^{(lin.)}(0)$ & 
$\lambda_0^{(lin.)}$ & $f_0^{(quad.)}(0)$ & $\lambda_0^{(quad.)}$ \\ \hline \hline
 $0.13390-0.13440 $ & $0.9981(3)$ & $2.5(3)$ & $0.9985(2)$ & $1.9(2)$ & $0.9983(3)$ & $2.2(4)$ \\ \hline
 $0.13390-0.13490 $ & $0.9912(16)$ & $2.9(4)$ & $0.9941(10)$ & $2.1(2)$ & $0.9915(15)$ & $2.8(4)$ \\ \hline
 $0.13390-0.13520 $ & $0.9858(41)$ & $3.2(6)$ & $0.9928(22)$ & $2.2(3)$ & $0.9858(40)$ & $3.2(6)$ \\ \hline
 $0.13440-0.13390 $ & $0.9981(2)$ & $2.4(3)$ & $0.9985(1)$ & $1.9(2)$ & $0.9984(2)$ & $2.1(3)$ \\ \hline
 $0.13440-0.13490 $ & $0.9975(5)$ & $3.3(5)$ & $0.9983(3)$ & $2.5(3)$ & $0.9976(5)$ & $3.2(5)$ \\ \hline
 $0.13440-0.13520 $ & $0.9940(21)$ & $3.7(7)$ & $0.9972(13)$ & $2.7(4)$ & $0.9946(19)$ & $3.5(7)$ \\ \hline
 $0.13490-0.13390 $ & $0.9923(11)$ & $2.5(3)$ & $0.9941(7)$ & $1.9(2)$ & $0.9931(11)$ & $2.3(3)$ \\ \hline
 $0.13490-0.13440 $ & $0.9978(4)$ & $2.9(4)$ & $0.9984(3)$ & $2.2(2)$ & $0.9980(4)$ & $2.7(4)$ \\ \hline
 $0.13490-0.13520 $ & $0.9993(4)$ & $4.3(9)$ & $0.9998(3)$ & $3.3(5)$ & $0.9996(4)$ & $3.6(9)$ \\ \hline
 $0.13520-0.13390 $ & $0.9882(22)$ & $2.5(3)$ & $0.9912(16)$ & $1.8(2)$ & $0.9894(21)$ & $2.2(3)$ \\ \hline
 $0.13520-0.13440 $ & $0.9951(13)$ & $2.9(4)$ & $0.9967(10)$ & $2.2(3)$ & $0.9955(11)$ & $2.7(4)$ \\ \hline
 $0.13520-0.13490 $ & $0.9993(3)$ & $3.9(7)$ & $0.9997(3)$ & $2.9(4)$ & $0.9995(3)$ & $3.4(7)$ \\ \hline \hline
 $0.13385-0.13433 $ & $0.9982(3)$ & $2.6(4)$ & $0.9986(2)$ & $2.0(2)$ & $0.9986(3)$ & $1.9(5)$ \\ \hline
 $0.13385-0.13501 $ & $0.9863(34)$ & $3.5(7)$ & $0.9913(17)$ & $2.6(3)$ & $0.9923(34)$ & $2.4(7)$ \\ \hline
 $0.13433-0.13385 $ & $0.9981(2)$ & $2.7(4)$ & $0.9985(1)$ & $2.0(2)$ & $0.9985(3)$ & $2.1(4)$ \\ \hline
 $0.13433-0.13501 $ & $0.9945(15)$ & $4.0(8)$ & $0.9966(9)$ & $2.9(4)$ & $0.9971(14)$ & $2.7(7)$ \\ \hline
 $0.13501-0.13385 $ & $0.9880(16)$ & $3.0(4)$ & $0.9913(9)$ & $2.2(2)$ & $0.9902(16)$ & $2.5(4)$ \\ \hline
 $0.13501-0.13433 $ & $0.9959(8)$ & $3.2(5)$ & $0.9972(5)$ & $2.4(3)$ & $0.9969(8)$ & $2.5(5)$ \\ \hline

\hline

\end{tabular}

\end{center}

\caption{\it The form factor at zero-momentum $f_0(0)$ and its slope $\lambda_0$, obtained 
from the three fits (\ref{eq:polar}-\ref{eq:quadratic}), for all combinations of the 
hopping parameters.}

\label{tab:f0(0)}

\end{table}

Before closing this Section we mention that our results for the slope $\lambda_0$, 
extrapolated to the physical meson masses using a linear dependence in the quark masses,
give in units of the physical value of $M_{\pi^+}^2$: $\lambda_0^{(pol.)} = 0.0122(22)$, 
$\lambda_0^{(lin.)} = 0.0089(11)$ and $\lambda_0^{(quad.)} = 0.0115(26)$. Our ``polar" 
value $\lambda_0^{(pol.)}$ is consistent with the recent determination from KTeV 
$\lambda_0 = 0.01414 \pm 0.00095$ \cite{KTEV_slope}, obtained using a pole parameterization.

\subsection{Discretization effects \label{subsec:discre}}

Lattice artifacts on $f_0(0)$ due to the finiteness of the lattice spacing start at 
$\cO(a^2)$ and are proportional to $(m_s - m_\ell)^2$, like the physical SU(3)-breaking 
effects. Indeed, the determination of $f_0(q_{\rm max}^2)$ is affected only by discretization
errors of $\cO[a^2 (m_s - m_\ell)^2]$, because the double ratio (\ref{eq:fnal1}) 
is $\cO(a)$-improved and symmetric with respect to the exchange $m_s \leftrightarrow m_\ell$ in the weak vertex. 
In addition, since $q_{\rm max}^2$ is proportional to $(m_s - m_\ell)^2$, $\cO(a^2)$ effects in 
the extrapolation from $q_{\rm max}^2$ to $q^2 = 0$ also vanish quadratically in 
$(m_s - m_\ell)$. Being in our calculation $a^{-1} \simeq 2.7$ GeV, we expect 
discretization errors to be sensibly smaller than the physical SU(3)-breaking effects. 

A qualitative estimate of discretization errors can be obtained by considering the effect of 
the terms proportional to the improvement coefficient $b_V$, which cancel out at first order in 
the double ratio (\ref{eq:fnal1}), but contribute at second order. A simple calculation 
shows that this correction is given by $[ 1 + b_V (a m_s + a m_\ell) / 2]^2 / [ (1 + b_v a m_s) 
\cdot (1 + b_V a m_\ell) ] \simeq 1 - a^2 b_V^2 (m_s - m_\ell)^2 / 4$. The size of such a $\cO(a^2)$ 
contribution is a few percent of the whole result. Calculations performed at different values of 
the lattice spacing, combined with the extrapolation to the continuum limit, will certainly allow 
a quantitative estimate of discretization errors and a further reduction of this source of uncertainty.

\begin{figure}[htb]

\begin{center}

\epsfxsize=11.5cm \epsffile{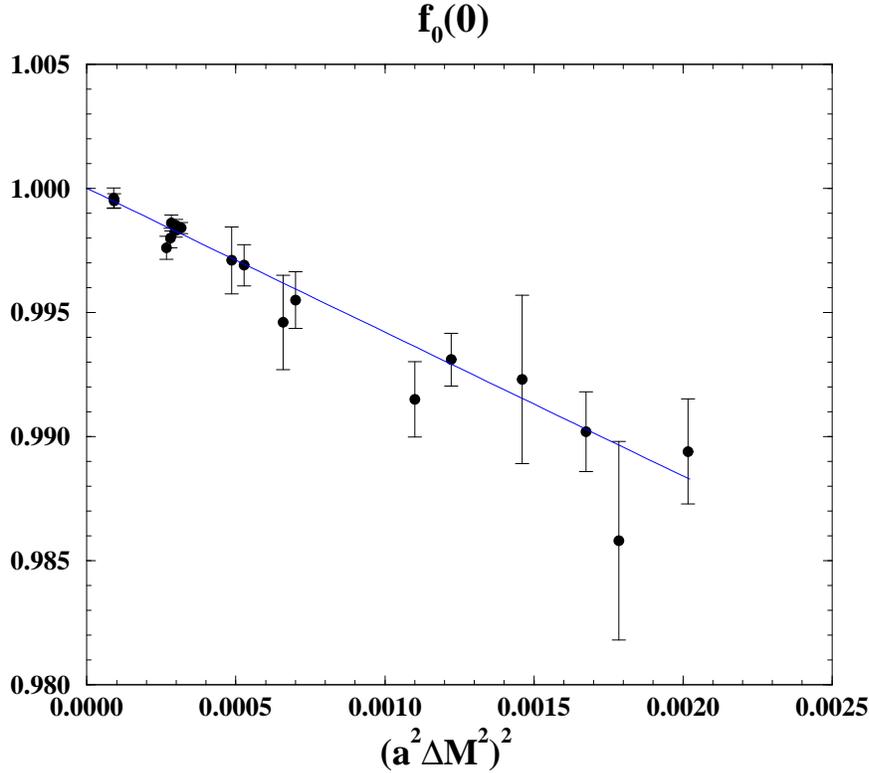} 

\end{center}

\caption{\it Values of $f_0(0)$, obtained from the quadratic
fit (\ref{eq:quadratic}), versus $(a^2 \Delta M^2)^2$. The solid line is the result of the
linear fit (\ref{eq:lin}).}

\label{fig:plotf0}

\end{figure}

In Fig.~\ref{fig:plotf0} we plot the values of $f_0(0)$, obtained from the
fit in $q^2$ given by Eq.~(\ref{eq:quadratic}), versus $(a^2 \Delta M^2)^2$. 
The results agree well with the quadratic dependence on $a^2 \Delta M^2$, expected from both
physical and lattice artifact contributions, as shown in
Fig.~\ref{fig:plotf0} by the solid line representing a na\"ive fit to the form 
\be
f_0(0) = 1 - A ~ (a^2 \Delta M^2)^2 ~ ,
\label{eq:lin}
\ee
where $A$ is a mass-independent parameter. 

\section{Extraction of $\Delta f$}
\label{sec:Df}

In order to determine the physical value of $f_0(0)$, we need 
to extrapolate our results of Table \ref{tab:f0(0)} to the physical kaon and pion masses. 
As discussed in the introduction, the problem of the 
chiral extrapolation is substantially simplified if we remove 
the effect of the leading chiral logs. 
To this purpose we consider the quantity
\be
\Delta f = 1  + f_2 -f_0(0) \,,
\label{eq:deltaf}
\ee
where $f_2$ represents the leading non-local contribution determined by pseudoscalar meson loops
within CHPT. This procedure is well defined thanks to the finiteness of the leading non-local 
contributions both in the quenched and in the unquenched version of the effective theory. 
Having subtracted the leading logarithmic corrections, we expect $\Delta f$ to receive large 
contributions from local operators in the effective theory and, as in the case of the $q^2$ slopes 
discussed in Section \ref{sec:extrq20}, to be better suited for a smooth polynomial
extrapolation in the meson masses. 

As pointed out in the Introduction, the subtraction of $f_2$ in Eq.~(\ref{eq:deltaf}) 
does not imply necessarily a good convergence of CHPT at order $\cO(p^4)$ for the meson masses 
used in our lattice simulations. The aim of this subtraction is to define a quantity 
whose chiral expansion starts at order $\cO(p^6)$ independently of the values of the meson masses, 
or a quantity better suited for a smooth chiral extrapolation. It is also worth to emphasise that 
the numerical impact of $f_2^q$ turns out to be very small (around or below 10 \% with respect to 
the whole value of $\Delta f$) for the highest meson masses used in our simulations. It increases 
(without exceeding 30 \% of $\Delta f$) in the case of lighter masses, where the chiral expansion 
should have a better convergence. 

\subsection{Chiral loops within full QCD}
In the isospin-symmetric limit, within full QCD, 
the expression of the leading chiral correction $f_2$ is \cite{LeuRo}
\begin{equation}
f_2={3\over 2} H_{\pi K} + {3 \over 2} H_{\eta K} \; ,
\label{eq:f2-full}
\end{equation}
where 
\begin{equation}
H_{PQ}= -{1 \over 64 \pi^2 f_\pi^2} \left[ M_P^2+M_Q^2+ {2 M_P^2 M_Q^2
    \over M_P^2 - M_Q^2} \ln {M_Q^2 \over M_P^2} \right] \; \; .
\label{eq:HH}
\end{equation}
Note that $f_2$ is completely specified in terms of pseudoscalar
meson masses and decay constants ($f_\pi \approx 132$ MeV); it is negative
($f_2 \approx -0.023$ for physical masses), as implied by unitarity \cite{LeuRo,FLRS}; it vanishes 
as $(M_K^2-M_\pi^2)^2/(f_\pi^2 M_K^2)$ in the SU(3) limit, 
following the combined constraints of chiral symmetry and the
Ademollo-Gatto theorem. 

\subsection{Chiral loops within quenched QCD}
\label{subs:chlogs}
The structure of chiral logarithms appearing in Eqs.~(\ref{eq:f2-full})--(\ref{eq:HH}),
is valid only in the full theory. In the quenched theory, the leading (unphysical) 
logarithms are instead those entering the one-loop functional of qCHPT 
\cite{cp-q,bg-q,Sharpe}. We calculated this correction and present the results in this Section. 
Normalizing the lowest-order qCHPT Lagrangian as in Ref.~\cite{cp-q}, with a quadratic term 
for the singlet field $\Phi_0={\rm str}(\Phi)$ chosen as
\beq
\left. \cL^q_2 \right\vert_{\Phi^2_0} = \frac{\alpha}{6} D_\mu \Phi_0 D^\mu \Phi_0 
- \frac{M_0^2}{6} \Phi_0^2~,
\eeq 
we find 
\beq
f_2^q  = H^q_{\pi K} +  H^q_{(s\bar{s}) K}~,
\label{eq:f2-quenched}
\eeq
where
\beq
 H^q_{P K} =  \frac{M_K^2}{96 \pi^2 f_\pi^2}\l [ 
  \frac{ M_0^2 ( M_K^2+M_{P}^2) - 2 \alpha M_K^2 M_{P}^2  }
  { {\left( M_K^2-M_{P}^2 \right) }^2}\,\log \l (\frac{M_{K}^2}{M_{P}^2}\r) -\,\alpha \r ]
\eeq
with $M_{(s\bar{s})}^2=2M_K^2-M_\pi^2$. As anticipated, the one-loop result in 
Eq.~(\ref{eq:f2-quenched}) is finite because of the
Ademollo-Gatto theorem, which is still valid in the quenched 
approximation \cite{cp-q} and forbids the appearance of contributions from local
operators in $f^q_2$. 
A proof beyond the one-loop level that the Ademollo-Gatto theorem 
(and more generally the Sirlin's relation \cite{s-sr}) holds within qCHPT 
can easily be obtained by applying the functional formalism to the 
demonstration  in Ref.~\cite{s-sr}. The latter needs only flavor symmetries 
which hold on the lattice also in the quenched case.

It is worth to emphasize that the nature of the SU(3) 
breaking corrections in the quenched theory is completely different 
from that of full QCD: only contributions coming from the mixing 
with the flavor-singlet state are present and one finds $f^q_2>0$, which is a 
signal of the non-unitarity of the theory. 
For typical values of the singlet parameters ($M_0 \approx 0.6$ GeV
and $\alpha \approx 0$ \cite{bardeen}) and for the physical values of pion and kaon masses,
one finds $f^q_2 \approx + 0.022$. For the quark masses used 
in the simulation (and at the same values of $M_0$ and $\alpha$),
the non-local correction is substantially smaller, $f^q_2 < 0.003$.
This estimate is not changed significantly if a value of $M_0$ as large as 
$1$ GeV and/or a $20 \%$ variation of $f_{\pi}$ are considered. We anticipate 
that the effect of the above-mentioned variations of $M_0$ and/or 
$f_{\pi}$ on our final result for $\Delta f$ is within $20 \%$ and it has been 
included in the quoted final uncertainties [see later Eq.~(\ref{eq:Deltaf_fin})].
The reason why in Fig.~\ref{fig:plotf0} all values of $f_0(0)$ are found to be 
smaller than unity, as expected in a unitary theory, may be that for these 
sets of masses the contribution of local operators dominates over the quenched 
chiral logs. 

The high level of accuracy required in the study of the $K_{\ell 3}$ form factors can only
be achieved once all possible sources of systematic errors are shown to be kept well under
control. In a lattice calculation this includes in particular a study
of finite volume effects. In our case such effects are expected to be smaller than
other systematic effects, as we have explicitly checked by performing the analytical calculation 
presented in the Appendix and based on the techniques discussed in Refs.~\cite{gl-fv,dg}.

\subsection{Extrapolation to the physical masses}
The evaluation of $f_2^q$ allows to express the lattice results in terms of the 
subtracted quantity $\Delta f$ defined in Eq.~(\ref{eq:deltaf}). To study the dependence of 
this quantity on the meson masses it is convenient to divide $\Delta f$ by $(\Delta M^2)^2$:
\beq
R(M_K, M_\pi) = \dfrac{\Delta f}{(a^2 \Delta M^2)^2} = 
\dfrac{ 1  + f^q_2(M_K, M_\pi) -f_0(0; M_K, M_\pi) }{(a^2 \Delta M^2)^2} ~.
\label{eq:linfit}
\eeq
We have used $f_2^q$ computed from Eq.~(\ref{eq:f2-quenched})
at the corresponding values of meson masses, setting $M_0 = 0.6\,$GeV
and $\alpha = 0.05$. The error induced by a variation of $M_0$ and $\alpha$ 
in a reasonable range of values is found to be negligible compared to the statistical error. 
We find that at the simulated masses the effect of $f_2^q$ does not exceed $30 \%$ of the value
of $[1 - f_0(0)]$. Since $f_2^q$ has a  non-trivial, non-analytic dependence from the quark 
masses, we subtract its contribution at each value of the quark masses.

\begin{figure}[htb]

\begin{center}

\epsfxsize=11.5cm \epsffile{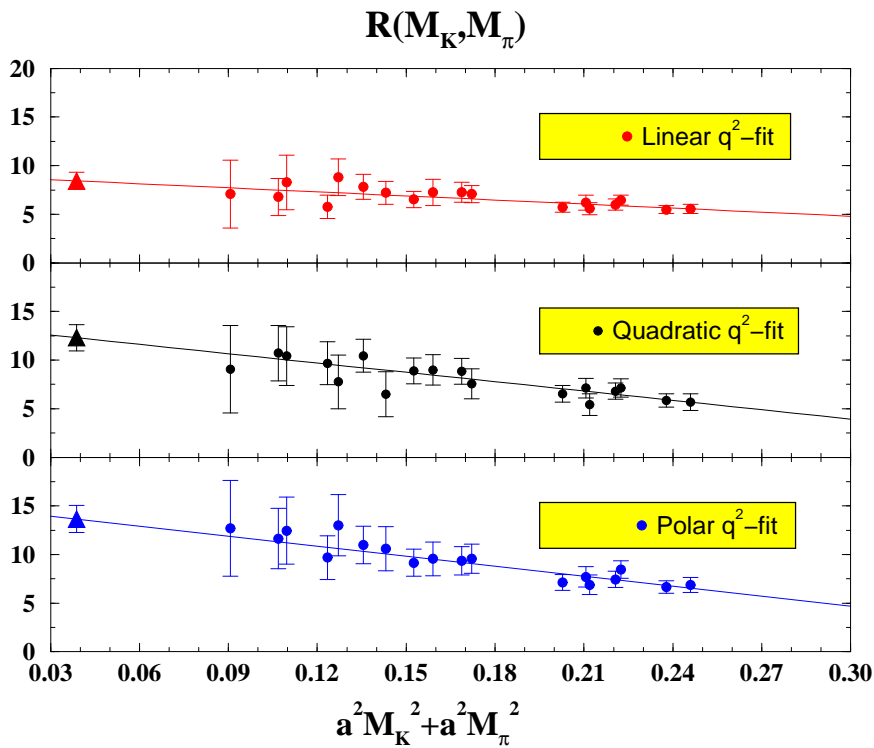}

\end{center}

\caption{\it $R(M_K, M_\pi)$ as a function of $[a^2 M_K^2 + a^2 M_{\pi}^2]$ for the cases of the 
linear, quadratic and polar $q^2$-extrapolation in the fit of $f_0(q^2)$. The solid lines represent 
the results of the linear fit (\protect\ref{eq:agfit}). Triangles indicate the values of $R(M_K, M_\pi)$ extrapolated to the
physical kaon and pion masses.}

\label{fig:linfit} 

\end{figure}

\begin{figure}[htb]

\begin{center}

\epsfxsize=11.5cm \epsffile{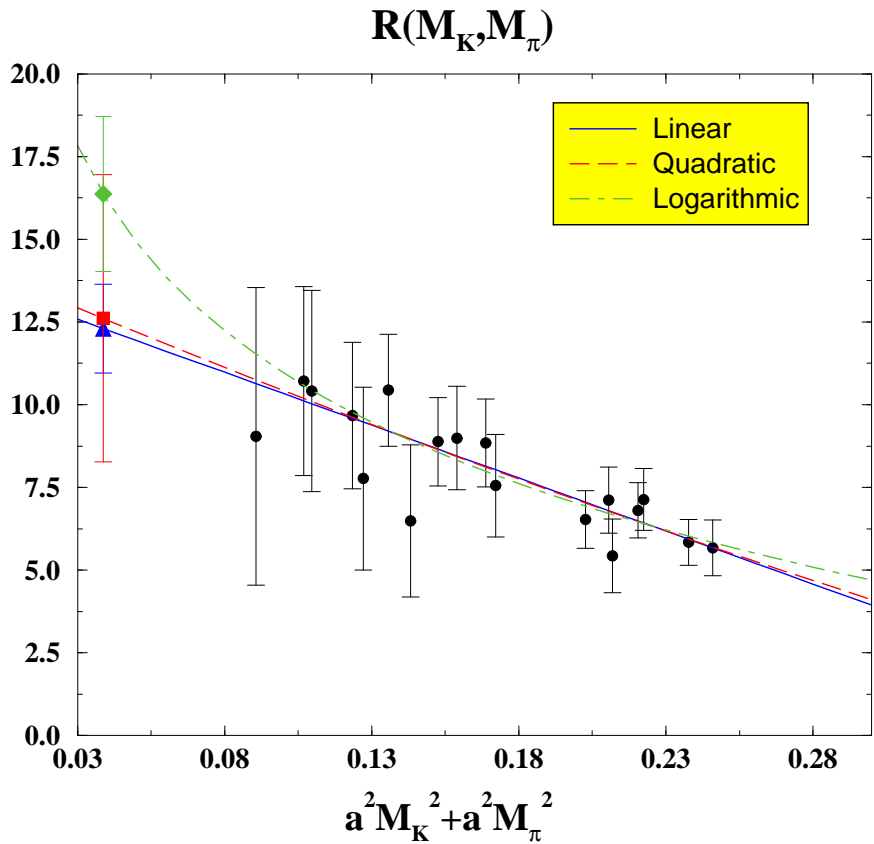} 

\end{center}

\caption{\it Comparison among linear (\ref{eq:agfit}), quadratic (\ref{eq:qmfit}) 
and logarithmic (\ref{eq:qlfit}) fits of the
ratio $R(M_K, M_\pi)$ as a function of
$[a^2 M_K^2 + a^2 M_{\pi}^2]$. Triangle, square and diamond
represent the values of $R(M_K, M_\pi)$ extrapolated to the physical meson masses.
For illustrative purposes we have chosen the case in which a quadratic $q^2$-dependence 
of $f_0(q^2)$ has been used to extrapolate the scalar form factor to $q^2 = 0$.}

\label{fig:fitcfr}

\end{figure}

As shown in Fig.~\ref{fig:linfit}, the dependence of $R(M_K, M_\pi)$ on the squared meson 
masses is well described by a simple linear fit:
\be
R(M_K, M_\pi) = c_{11} + c_{12} [(a M_K)^2 + (a M_\pi)^2] ~ ,
\label{eq:agfit}
\ee
whereas the dependence on $\Delta M^2$ is found to be negligible.
We find an excellent $\chi^2$ for all the three sets of values of $f_0(0)$
obtained from the linear, quadratic and polar extrapolation in $q^2$, see 
Fig.~\ref{fig:linfit}. 
In order to check the stability of the results, we have also 
performed quadratic and logarithmic fits, viz.
\bea
\label{eq:qmfit}
R(M_K, M_\pi) & = & c_{21} + c_{22} [(a M_K)^2 + (aM_\pi)^2] + c_{23} [(a M_K)^2 + (a M_\pi)^2]^2\, , \\
R(M_K, M_\pi) & = & c_{31} + c_{32} \log[ (a M_K)^2 + (a M_\pi)^2]\, .
\label{eq:qlfit}
\eea
It is reassuring to find that, while increasing the number of parameters 
in the chiral extrapolation as done in the quadratic fit (\ref{eq:qmfit}) 
leads to larger uncertainties, the shift in the central values remains smaller than 
the errors. The logarithmic fit has been considered in order to
investigate the presence of possible important non-analytical terms in 
$R(M_K, M_\pi)$\footnote{We have also considered other fitting procedures, like a linear
fit applied to the subset of data corresponding to $[(a M_K)^2 + (aM_\pi)^2] < 
0.18$. The extrapolated value of $R(M_K, M_\pi)$ at the physical masses is well 
within the spread of values obtained using the linear, quadratic and logarithmic 
fits.}. In Fig.~\ref{fig:fitcfr} we show that linear, quadratic and logarithmic
functional forms provide equally good fits of the data with consistent 
results for the extrapolation to the physical point.

\begin{table}[htb]

\vspace{0.5cm}

\begin{center}

\begin{tabular}{||l||c|c|c||}
\hline
 & Linear Fit & Quadratic Fit  & Logarithmic Fit  \\
\hline
Linear $f_0(q^2)$ & $0.009 \pm 0.001$ & $0.008 \pm 0.003$ & 
$0.011 \pm 0.002$\\ 
\hline
Quadratic $f_0(q^2)$ & $0.013 \pm 0.002$ & $0.014 \pm 0.005$ & 
$0.018 \pm 0.003$\\  
\hline
Polar $f_0(q^2)$ & $0.015 \pm 0.002$ & $0.018 \pm 0.005$ & 
$0.020 \pm 0.003$\\
\hline
\end{tabular}

\vspace{0.5cm}

\begin{tabular}{||l||c|c|c||}
\hline
 & Linear Fit & Quadratic Fit  & Logarithmic Fit  \\
\hline
Linear $f_0(q^2)$ & $0.012 \pm 0.001$ & $0.013 \pm 0.003$ & 
$0.015 \pm 0.002$\\ 
\hline
Quadratic $f_0(q^2)$ & $0.017 \pm 0.002$ & $0.021 \pm 0.004$ & 
$0.023 \pm 0.002$\\  
\hline
Polar $f_0(q^2)$ & $0.019 \pm 0.002$ & $0.025 \pm 0.005$ & 
$0.026 \pm 0.003$\\
\hline
\end{tabular}

\end{center}

\caption{\it Results for $\Delta f$ obtained from linear, quadratic and
logarithmic fits of the ratio $R(M_K, M_\pi)$ assuming linear, 
quadratic or polar functional forms for the extrapolation of the scalar form factor 
to $q^2 = 0$. Upper and lower tables correspond to different choices of the time
interval chosen for the fits of the two-point correlation functions (see text).}

\label{tab:Deltaf}

\end{table}

The extrapolated values of $R(M_K, M_\pi)$ are converted into predictions 
for $\Delta f$ by using the values of the physical kaon and pion masses in lattice 
units given in Section~\ref{sec:basics}. By this procedure, i.e.~expressing both 
$R(M_K, M_\pi)$ and the meson masses in lattice units, we get rid of the error 
associated with the determination of the lattice spacing in the quenched approximation. 
In Table~\ref{tab:Deltaf} we collect the values of $\Delta f$ obtained for all 
the functional forms assumed both in fitting the $q^2$-dependence of $f_0(q^2)$ 
and in extrapolating the ratio $R(M_K, M_\pi)$ to the physical meson masses. 
We present two sets of results. The first set, given in the upper table, is obtained by 
determining the matrix elements $\sqrt{Z_K}$ and $\sqrt{Z_\pi}$ [see Eq.~(\ref{eq:c2ptexp})] 
from a fit of the two-point correlation functions in the same time interval chosen for the 
three-point correlators, namely $t/a \in$ [11,17]. For the second set, given in the lower table, 
the same time interval chosen to extract the meson masses, i.e.~$t/a \in$ [12,26], has been 
considered.

From the spread of the values in Table~\ref{tab:Deltaf} we quote our final estimate
\be
\Delta f =  R(M_K^{\rm phys}, M_\pi^{\rm phys}) \times [(a^2 \Delta M^2)^2]^{\rm phys} =
 (0.017 \pm 0.005_{\rm stat} \pm 0.007_{\rm syst} ) ~,
\label{eq:Deltaf_fin}
\ee
where the systematic error is dominated by the extrapolation of $f_0(q^2)$ in $q^2$ and 
of $R(M_K, M_\pi)$ in the meson masses.

The result in Eq.~(\ref{eq:Deltaf_fin}) is the main result of this paper. We have 
checked that all other sources of systematic errors, as those originated
from different choices of the plateau intervals, from the uncertainties in the
values of the renormalization constant $Z_V$ and of the improvement coefficients
$b_V$ and $c_V$, and from the estimated size of discretization effects, are 
negligible with respect to the systematic error quoted in Eq.~(\ref{eq:Deltaf_fin}).

\section{Estimate of $f_+(0)$ and impact on $|V_{us}|$}
The value of $\Delta f$ given in Eq.~(\ref{eq:Deltaf_fin}) is in excellent agreement with 
the estimate of $\Delta f$ made by Leutwyler and Roos
using a general parameterization of the SU(3) breaking structure 
of the pseudoscalar-meson  wave functions \cite{LeuRo}. 
It must be stressed, however, that our result is the first 
computation of this quantity using a non-perturbative method 
based on QCD. Combining our estimate of $\Delta f$ with the physical value of $f_2$ 
from Eq.~(\ref{eq:f2-full}), we finally obtain
\be
 f_+^{K^0\pi^-}(0) = 0.960 \pm 0.005_{\rm stat} \pm 0.007_{\rm syst}
\label{eq:f0final}
\ee
to be compared with the value $f_+^{K^0\pi^-}(0) = 0.961 \pm 0.008$ 
given in Ref.~\cite{CKMBook} and quoted by the PDG \cite{PDG}. 

\subsection{Comparison with other approaches}
We now compare our results in Eqs.~(\ref{eq:Deltaf_fin}-\ref{eq:f0final})
with the recent evaluations of the $\cO(p^6)$ 
contributions to $f_+(0)$ made in Refs.~\cite{BT,CKNRT2,Jamin}.
In the chiral expansion the whole $\cO(p^6)$ term can be separated 
into two terms, coming respectively from meson loops and local terms. Whereas the whole
$\cO(p^6)$ term is scale- and scheme-independent, the two terms separately depend on the
renormalization scale and scheme. Leutwyler-Roos assumed that their quark-model
estimate represents the whole physical $\cO(p^6)$ term. In Refs.~\cite{BT,CKNRT2,Jamin} this
result was interpreted instead to represent the contribution due to the local terms only, at
the renormalization scale of the $\rho$-meson mass. 
The full $\cO(p^6)$ contribution was then obtained by adding 
the renormalized loop amplitude computed in Ref.~\cite{BT}. In this way
the whole $\cO(p^6)$ correction is found to be very small, though within a large error.
The compensation between local and non-local terms at
$\cO(p^6)$ strongly depends on the choice of the renormalization scale.
Indeed, varying the renormalization 
scale from $\mu_{\overline{\rm MS}} \approx 1$ GeV 
to $\mu_{\overline{\rm MS}} \approx 0.5$ GeV, 
the loop contribution to $f_+(0)$ changes from $+0.4\%$ 
to $+3.5\%$ \cite{CKNRT2}.

As pointed out in Ref.~\cite{Jamin}, the use of dispersion relations could help to 
solve the scale ambiguity. 
By means of dispersion relations one obtains complementary 
constraints on the slope and curvature of $f_0(q^2)$ that, in turn,
can be used in conjunction with the chiral 
calculation of Ref.~\cite{BT} to determine $f_+(0)$. 
Present data, however, are not accurate enough: the estimate 
of $\cO(p^6)$ local terms given in Ref.~\cite{Jamin}
can only be interpreted as a consistency check of the 
Leutwyler-Roos ansatz, rather than as a truly independent 
determination. Finally, it must be stressed that 
all the purely analytic approaches to the estimate of $f_+(0)$ suffer 
from an intrinsic uncertainty due to the value of 
$f_K/f_\pi$ \cite{Fuchs}.

Both these ambiguities (the scale uncertainties and the 
dependence on $f_K/f_\pi$) are not present in our approach.
By construction, $\Delta f$ is a scale-independent quantity 
which includes all contributions beyond $\cO(p^4)$.
We determine this quantity only in terms of meson masses.
The main sources of uncertainty in our approach come from 
the $q^2$-dependence of the form factor, the chiral extrapolation and 
the quenched approximation, 
but we argue that such uncertainties are included in 
the systematic error quoted in Eq.~(\ref{eq:Deltaf_fin}).

\subsection{Impact on $|V_{us}|$}
Given the good agreement between our estimate 
of $f_+^{K^0\pi^-}(0)$ and the value of Refs.~\cite{CKMBook,PDG}, our result implies 
a negligible shift in the determination of $|V_{us}|$ 
from $K_{\ell 3}$ with respect to those works. 
Assuming the same experimental inputs, the value 
computed with our form factor, applying
the updated radiative and isospin-breaking corrections of Refs.~\cite{CKNRT2,CKNRT},
is $|V_{us}| = 0.2202 \pm 0.0025$ to be compared with 
$|V_{us}| = 0.2196 \pm 0.0026$ of Refs.~\cite{CKMBook,PDG}.
As pointed out by several authors, this estimate shows a $\approx 2 \sigma$ 
deviation from the CKM unitarity relation, which yields $|V_{us}| = 0.2278 
\pm 0.0020$, taking into account the recent determination $|V_{ud}| = 0.9737 
\pm 0.0007$ from Ref.~\cite{Vud}. According to our analysis and assuming that 
the quenching effects do not drastically change our findings, this deviation 
cannot be attributed to an underestimate of the amount of SU(3) breaking 
in $f_+^{K^0\pi^-}(0)$. 

We stress, however, that the above estimate of $|V_{us}|$ takes into account 
only the (old) published $K_{\ell 3}$ data. Recently, new results have been 
presented \cite{E865,KTEV,KLOE,NA48}. For instance, using only the high-statistics 
$K_{e3}$ results of Refs.~\cite{E865} and \cite{KTEV}, we find substantially 
higher values, namely $|V_{us}| = 0.2275 \pm 0.0030$ and $|V_{us}| = 0.2255 
\pm 0.0025$, respectively, which are in good agreement with CKM unitarity.
The KTeV findings \cite{KTEV} are also confirmed by the preliminary results 
presented by KLOE \cite{KLOE} and NA48 \cite{NA48} at the ICHEP '04 Conference.

\section{Conclusions}

We presented a quenched lattice study of the $K \to \pi$ vector form factor at zero-momentum
transfer. Our calculation is the first one obtained by using a non-perturbative method 
based only on QCD, except for the quenched approximation.
Our main goal is the determination of the second-order SU(3)-breaking quantity $[1 - f_+(0)]$,
which is necessary to extract $|V_{us}|$ from $K_{\ell 3}$ decays. In order to reach the
required level of precision we employed the double ratio method originally proposed
in Ref.~\cite{FNAL} for the study of heavy-light form factors. We 
found that this approach allows to calculate the scalar form factor $f_0(q^2)$ at 
$q^2 = q_{\rm max}^2$ with a statistical uncertainty well below $1 \%$.

A second crucial ingredient is the extrapolation of the scalar form factor to $q^2 = 0$. 
This was performed by fitting accurate results obtained using suitable double ratios of three-point
correlation functions. The systematic error due to finite volume effects was evaluated and 
discretization errors were qualitatively estimated to be a few percent of the deviation of $f_0(0)$ 
from unity. Calculations performed at different values of the lattice spacing, combined with the 
extrapolation to the continuum limit, will certainly allow a more quantitative estimate of discretization 
errors and a further reduction of this source of uncertainty. Nevertheless it is reasonable to conclude 
that the uncertainties coming from the functional dependencies of the scalar form factor on both 
$q^2$ and the meson masses are the dominant contribution to the systematic error.

The leading chiral artifacts of the quenched approximation, $f_2^q$, were corrected 
for by means of an analytic calculation in quenched chiral perturbation theory. 
After this subtraction, the lattice results have been smoothly extrapolated to 
the physical meson masses. Our final value for $f_+^{K^0\pi^-}(0)$ is
\be
 f_+^{K^0\pi^-}(0) = 0.960 \pm 0.005_{\rm stat} \pm 0.007_{\rm syst} 
 = 0.960 \pm 0.009 ~ ,
\label{eq:final}
\ee
where the systematic error does not include an estimate of quenched effects beyond 
$\cO(p^4)$.

The impact of our result on the determination of $|V_{us}|$ was also 
addressed. Using the (old) published $K_{\ell 3}$ data from 
Ref.~\cite{PDG}, we obtain $|V_{us}| = 0.2202 \pm 0.0025$, which still implies a $\approx 2 \sigma$ 
deviation from the CKM unitarity relation. We stress that, according to our 
analysis, such a deviation should not be attributed to an 
underestimate of the amount of SU(3) breaking effects in $f_+^{K^0\pi^-}(0)$.
Using only the recent high-statistics $K_{e3}$
results of Refs.~\cite{E865,KTEV} one finds substantially 
higher values, $|V_{us}| = 0.2275 \pm 0.0030$ and $|V_{us}| = 0.2255 \pm 0.0025$, 
which are in good agreement with CKM unitarity.

In order to reach a precision better than $1 \%$ in the determination of 
the Cabibbo angle both theoretical and experimental improvements are needed.
As for the experimental side, new high-statistics results on both charged and neutral 
$K_{\ell 3}$ modes are called for. As for the theoretical side, the next important steps are: 
i) to remove the quenched approximation; ii) to decrease the values of the simulated 
meson masses in order to gain a better control over the chiral 
extrapolation of lattice results, and iii) to use larger lattice volumes for decreasing
lattice momenta, in order to improve the determination of the slope of the scalar
form factor.

\section*{Acknowledgments}
The work of G.I.~and F.M.~is partially supported by IHP-RTN,
EC contract No.~HPRN-CT-2002-00311 (EURIDICE).

\section*{Appendix}

In this Appendix we present the results of the analytic calculation of finite volume corrections
to $f_2$, both in the quenched and unquenched theory.
 
An important advantage of the effective theory approach is that it allows to evaluate 
(and eventually to correct for) the lattice artifacts due to a 
finite volume. This is simply achieved by imposing periodic boundary conditions 
to the wave functions of the chiral fields.
This approximation represents a good description of the finite-volume effects 
as long as the size $L$ of the lattice satisfies the condition $L \cdot M_{\pi,K} \gg 1$ \cite{gl-fv},
which is well satisfied in the present simulation.

Defining the finite-volume shifts as
\beq
\delta_L= \l. f_2\r|_L  - f_2\,,
\eeq
we find that the corrections to Eqs.~(\ref{eq:f2-full}) and (\ref{eq:f2-quenched}) are given by
\bea 
\delta_L&=&-\frac{3}{8\, f_\pi^2}\l[ 2\,\xi_\frac12(L,M_K) + \xi_\frac12(L,M_\pi) + 
      \xi_\frac12(L,M_\eta)  \r.\nn \\
       & & \qquad \l.- 2\,\xi_\frac12(L,M_\pi,M_K) - 2\,\xi_\frac12(L,M_\eta,M_K) \r]  \,, \label{eq:dL-full}\\
\delta_L^q&=&\frac{1}{72\, f_\pi^2}\l[-2\,\frac{\left( 5\,M_K^2\,\alpha  - 4\,M_\pi^2\,\alpha  - M_0^2 \right) }{M_K^2 - M_\pi^2} 
                \, \left( \xi_\frac12(L,M_\pi) - \xi_\frac12(L,M_{(s \bar{s})}) \right) \r.\nn \\
        &&\qquad \l.+ 3\,\left( M_0^2 - \alpha \,M_\pi^2\right) \,\xi_\frac32(L,M_\pi) + 
    3\,\left( M_0^2 -\alpha\,M_{(s \bar{s})}^2\right) \,\xi_\frac32(L,M_{(s \bar{s})}) \r.\nn \\
        &&\qquad \l. - 12\,\frac{\left( M_K^2\,\alpha  - M_0^2 \right)}{M_K^2 - M_\pi^2}  \,
       \left( \xi_\frac12(L,M_\pi,M_K) - \xi_\frac12(L,M_{(s \bar{s})},M_K) \right) \r.\nn \\
        &&\qquad \l.+  8\,\left( M_{(s \bar{s})}^2\,\alpha  - M_0^2 \right) \, \xi_\frac32(L,M_\pi,M_{(s \bar{s})}) 
                +12\,\left( M_\pi^2\,\alpha  - M_0^2 \right) \,\xi^\prime_\frac12(L,M_K,M_\pi)\r.\nn \\
        &&\qquad \l. + 12\,\left( M_{(s \bar{s})}^2\,\alpha  - M_0^2 \right) \, \xi^\prime_\frac12(L,M_K,M_{(s \bar{s})})\r]\,, \label{eq:dL-quenched}
\eea
in the unquenched and quenched cases, respectively. The functions $\xi_s$ are defined as
\bea
\xi_s(L,M_1) = \xi_s(L,M_1,M_1)
&=& \frac{1}{\Gamma(s)}\int_0^\infty d\tau \,\tau^{s-1}e^{- M_1^2 \tau} 
        \l[ \frac{1}{L^3}\theta^3\l(\frac{4\pi^2}{L^2}\tau\r)-\frac{1}{(4\pi\tau)^{3/2}}\r] \,, \nn \\
\xi_s(L,M_1,M_2)&=&\frac{1}{\Gamma(s)}\int_0^\infty d\tau \,\tau^{s-1}e^{-\frac{M_1^2+M_2^2}{2}\,\tau} 
        \l[\frac{\sinh\l(\frac{M_1^2-M_2^2}{2}\,\tau \r)}{\l(\frac{M_1^2-M_2^2}{2}\,\tau\r)}\r]\, 
        \nn \\  && \times  
\l[ \frac{1}{L^3}\theta^3\l(\frac{4\pi^2}{L^2}\tau\r)-\frac{1}{(4\pi\tau)^{3/2}}\r] \,,\nn \\
\xi^\prime_s(L,M_1,M_2)&=&-\frac{\partial}{\partial M_2^2}\xi_s(L,M_1,M_2)\,,
\eea
where $\theta(x) \equiv \sum_{n=-\infty}^{\infty} e^{-x\, n^2}$, and they
represent the differences between finite volume sums and infinite volume integrals 
(see Ref.~\cite{dg} for more details).
While $\delta_L$ represents a small correction in the range of masses used in the simulation,
the effect of $\delta_L^q$ is not negligible with respect to the size 
of the quenched logs in Eq.~(\ref{eq:f2-quenched}). Nevertheless, given the smallness of the 
leading chiral corrections at the simulated meson masses, we find that the inclusion of 
finite volume corrections has a negligible effect on our final estimate of $\Delta f$.

\end{document}